\documentclass[12pt]{spieman}  

\usepackage{graphicx} 
\usepackage{booktabs}
\usepackage[suffix=]{epstopdf}
\usepackage{setspace}
\usepackage{tocloft}

\usepackage{amsmath}  
\usepackage{amssymb}  

\usepackage{array}
\newcolumntype{L}[1]{>{\raggedright\let\newline\\\arraybackslash\hspace{0pt}}m{#1}}
\newcolumntype{C}[1]{>{\centering\let\newline\\\arraybackslash\hspace{0pt}}m{#1}}
\newcolumntype{R}[1]{>{\raggedleft\let\newline\\\arraybackslash\hspace{0pt}}m{#1}}

\newcommand{\R}{\mathcal R}
\newcommand{\I}{\mathcal I}

\title{Recursive Starlight and Bias Estimation for High-Contrast Imaging with an Extended Kalman Filter}
\author{A J Eldorado Riggs\supscr{a}, N. Jeremy Kasdin\supscr{a}, and Tyler D. Groff\supscr{a}}
\affiliation{\supscrsm{a}Princeton University, Department of Mechanical and Aerospace Engineering, Engineering Quadrangle, Princeton, NJ 08544, USA}

\begin{document}
\maketitle

\begin{abstract} 
For imaging faint exoplanets and disks, a coronagraph-equipped observatory needs focal plane wavefront correction to recover high contrast. The most efficient correction methods iteratively estimate the stellar electric field and suppress it with active optics. The estimation requires several images from the science camera per iteration. To maximize the science yield, it is desirable both to have fast wavefront correction and to utilize all the correction images for science target detection. Exoplanets and disks are incoherent with their stars, so a nonlinear estimator is required to estimate both the incoherent intensity and the stellar electric field. Such techniques assume a high level of stability found only on space-based observatories and possibly ground-based telescopes with extreme adaptive optics. In this paper, we implement a nonlinear estimator, the iterated extended Kalman filter (IEKF), to enable fast wavefront correction and a recursive, nearly-optimal estimate of the incoherent light. In Princeton's High Contrast Imaging Laboratory we demonstrate that the IEKF allows wavefront correction at least as fast as with a Kalman filter and provides the most accurate detection of a faint companion. The nonlinear IEKF formalism allows us to pursue other strategies such as parameter estimation to improve wavefront correction.
\end{abstract}

\keywords{wavefront sensing, wavefront control, high-contrast imaging, coronagraph, exoplanet}

{\noindent \footnotesize{\bf Address all correspondence to}: A J Eldorado Riggs; E-mail:  \linkable{ajriggs@jpl.nasa.gov}.  Published by JATIS with free access at doi:10.1117/1.JATIS.2.1.011017}


\begin{spacing}{1}   

\section{Introduction}\label{sec:intro}

Radial velocity and transit surveys have transformed our understanding of the universe by detecting thousands of planets outside our solar system. Dozens of these known exoplanets are close enough to image directly, which would allow us to obtain their spectra and fully determine their orbital parameters. Direct imaging requires high contrast in the image, factors of $10^{10}$ or more for Earth-size planets or $10^{9}$ for Neptune and Jupiter analogs. Atmospheric turbulence precludes obtaining such high contrast from the ground, so a space-based observatory is necessary. The proposed Coronagraph Instrument (CGI) on the Wide-Field Infrared Survey Telescope-Astrophysics Focused Telescope Asset (WFIRST-AFTA) is expected to image about 16 known cool gas-giant exoplanets and spectrally characterize about 6 of them.\cite{traub2015yield}  

A coronagraph uses a series of apodizers, masks, and stops in the optical train to modify or remove the Point Spread Function (PSF) of the telescope  and create image plane regions of high contrast where an exoplanet can be seen.  The optics for a coronagraph cannot be manufactured to the smoothness and reflectivity requirements to obtain $10^{-9}$ or better planet-to-star contrast passively.\cite{shaklan2006reflectivity}  A set of deformable mirrors (DMs) is necessary to mitigate these aberrations and recover a high-contrast region called a dark hole.  The CGI on WFIRST-AFTA will be equipped with two coronagraph types and two DMs, making it the first high-contrast coronagraphic mission in space with high-actuator-count wavefront control.

Wavefront correction for high-contrast coronagraphy differs from the method regularly used in astronomy. In traditional adaptive optics, the wavefront phase is measured at a pupil and conjugated by a DM.  That approach is inadequate for generating high contrast; a non-flat pupil phase is acceptable as long as the starlight destructively interferes in the image. In addition, uncontrollable, high-spatial frequencies in the pupil phase can mix and move light into the dark hole even if all correctable spatial frequencies are eliminated by the DMs. Finally, amplitude aberrations degrade contrast and cannot be mitigated with just phase conjugation. As a result, the wavefront control approach being planned for extreme high contrast in space relies on correction in the focal plane.  This requires forming an estimate of the focal plane electric field and then finding a DM command to improve the contrast.

The main challenge of focal plane wavefront correction with a high-contrast coronagraph is to sense the wavefront quickly. The most efficient correction routines need an estimate of the stellar electric field. A wavefront sensor cannot be used (alone) because it estimates only the phase and introduces non-common-path aberrations. The science camera is the only common-path sensor available, but exposure times can become very long as the contrast gets higher and the signal becomes fainter.  

Several techniques, all of which require one or more extra images for the estimator, exist for creating focal plane intensity diversity to calculate the electric field. The self-coherent camera (SCC)\cite{mazoyer2014scc} is an estimation technique for coronagraphs with focal plane phase masks. Pinholes outside the nominal beam radius at the Lyot stop produce an interference pattern in the image that is used to calculate the electric field. Another technique is COronagraphic Focal-plane wave-Front Estimation for Exoplanet detection (COFFEE),\cite{paul2014coffee} which utilizes a maximum a posteriori approach to estimate pupil plane aberrations and the bias signal. COFFEE introduces large phase aberrations at the DM to create diversity in the image plane. In this paper, we use the most tested and widely applicable estimation technique in which small, pair-wise probes are actuated on the DM to create image plane diversity.\cite{giveon2007broadband} 

High-contrast wavefront correction is an iterative process. With each electric field estimate, the controller suppresses as much starlight as possible in the dark hole. The contrast is then re-measured and more correction iterations are used until sufficiently high contrast is reached. Our prior work utilized a Kalman filter (KF) to estimate the stellar electric field recursively during wavefront correction. In this paper, we explore the use of a nonlinear filter, the extended Kalman filter (EKF), to estimate recursively both the stellar electric field and the intensity bias during wavefront correction. Science targets such as exoplanets and disks will be incoherent with a star, so they will appear in the bias estimate. Therefore, nonlinear, recursive wavefront correction lets us build the best possible real-time estimate of our potential science targets while recovering high contrast. The KF and EKF formalisms used in this paper are equally applicable to the SCC, and COFFEE could be easily modified to allow recursive estimation.

The recursive estimation techniques in this paper are discussed in the context of space-based observatories but may also apply to some ground telescopes, in particular those with extreme adaptive optics. If an observatory and the wavefront are stable enough for focal plane wavefront correction to function, then Kalman filtering should still be able to improve wavefront estimation by accounting for the model uncertainty of the system. Nonlinear estimation of the wavefront and bias is less robust to large uncertainties, however, and is most likely better suited for ultra-stable space telescopes.
 
\section{Review of Focal Plane Wavefront Correction}

In this section we describe the current progress in focal plane wavefront estimation and control. A longer discussion can be found in the paper by Groff et al.\ \cite{groff2015methods} in this issue. We re-derive important results to pose the problem and to establish the mathematical framework for the EKF derivation in Section \ref{sec:recursive}. 

\subsection{Linear Focal Plane Wavefront Control} \label{ControlOverview}	

The first successful controller for focal plane wavefront correction was speckle nulling.\cite{borde2006speckle} In this estimation-free scheme, sinusoids with different phases are applied to the DM to suppress stellar speckles at the targeted spatial frequency. Speckle nulling requires many hundreds or thousands of correction iterations, too many for use in a space mission. 

Model-based estimation and control enables faster correction. The first model-based controller was Electric Field Conjugation (EFC).\cite{giveon2007broadband} EFC minimizes energy in the dark hole and uses Tikhonov regularization to prevent too large of a command from being sent to the DMs. Because the electric field varies with wavelength, field estimates at several wavelengths within a larger bandpass are required to achieve broadband correction.  An alternative model-based controller, stroke minimization,\cite{pueyo2009optimal} minimizes DM actuation subject to a constraint on contrast. Stroke minimization has the same mathematical formula as EFC but provides a logical means of choosing the actuator regularization value at each correction iteration.\cite{groff2015methods} 

Here we derive the linearized electric field at the DM for use with the controller and estimator. Let $\tilde{E}_{0}(x,y)$ be the initial complex electric field at the DM including the incident field and the nominal complex aberrations on the DM, where $(x,y)$ are coordinates in the plane of the DM. Let $\phi_{k-1}(x,y)$ be the total phase contribution of the DM at correction iteration $k{-}1$, and let $\Delta \phi_{k}(x,y)$ be the perturbation of the DM phase at correction iteration $k$ such that
\begin{align}
\phi_{k-1}(x,y) = \sum^{k-1}_{j=1}\Delta \phi_{j}(x,y).
\end{align}
The phase at the DM is twice the surface height of the DM and scales inversely with wavelength $\lambda$ (in meters). Since for small deformations we can approximate the DM surface as the sum of the normalized actuator influence function $f(x,y)$ times the displacement command $\Delta u_{k,q}$ (in meters) at each actuator $q$'s center location $(x_q,y_q)$, the perturbation phase at the DM is given by
\begin{equation}
\Delta \phi_{k}(x,y) =  \frac{2}{\lambda} \sum_q^{N_{act}} \Delta u_{k,q} f(x - x_q, y - y_q),
\end{equation}
where $N_{act}$ is the number of DM actuators. 


Assuming small perturbation commands to the DM, we can approximate the electric field leaving the DM, $\tilde{E}_{k}(x,y)$, with a first order Taylor series expansion about the most recent DM perturbation,
\begin{align}
\tilde{E}_{k}(x,y) &= \tilde{E}_{0}(x,y) e^{i ( \phi_{k-1}(x,y) + \Delta \phi_{k}(x,y) )} \\
&\approx \tilde{E}_{0}(x,y) e^{i \phi_{k-1}(x,y) } \bigl(1 + i \Delta \phi_k (x,y)\bigr) ,
\label{eq:EpupDMTaylor}
\end{align}
Nearly all coronagraphs can be modeled as a series of linear operators such as Fourier transforms, Fresnel propagations, and mask multiplications. Since the control and estimation methods presented here are general to all coronagraphs, we represent the propagation from the DM to the science camera by the linear operator $\mathcal{C}\lbrace \cdot \rbrace$ to obtain the focal plane electric field $E_k(\xi,\eta)$,
\begin{align}
E_k(\xi,\eta) &=  \mathcal{C}\lbrace \tilde{E}_k(x,y) \rbrace \nonumber \\
	&\approx \mathcal{C}\lbrace \tilde{E}_{0}(x,y) e^{i \phi_{k-1}(x,y) } \rbrace + \mathcal{C}\lbrace i \tilde{E}_{0}(x,y) e^{i \phi_{k-1}(x,y) } \Delta \phi_k (x,y) \rbrace \nonumber \\
          &= E_{k-1}(\xi,\eta) + \sum_q^{N_{act}}  \Delta u_{k,q} \mathcal{C}\lbrace i \tilde{E}_{0}(x,y) e^{i \phi_{k-1}(x,y) } f(x - x_q,y-y_q) \rbrace \nonumber \\
          &= E_{k-1}(\xi,\eta) + \sum_q^{N_{act}}  \Delta u_{k,q} B_{k-1,q}(\xi, \eta),
\label{eq:EfocDMTaylor}          
\end{align}
where $(\xi, \eta)$ are coordinates in the image. The aberrated focal plane electric field before the new command at correction iteration $k$ is 
\begin{equation}
E_{k-1}(\xi,\eta)=\mathcal{C}\lbrace \tilde{E}_{0}(x,y) e^{i \phi_{k-1}(x,y) } \rbrace,
\end{equation}
and the Jacobian of each actuator at the image is given by the function
\begin{equation}
B_{k-1,q}(\xi, \eta) = \mathcal{C}\lbrace i \tilde{E}_{0}(x,y) e^{i \phi_{k-1}(x,y) } f(x - x_q,y-y_q) \rbrace.
\end{equation}

Detectors measure the intensity in finite-sized pixels, so the focal plane field is converted from the continuous coordinates  $(\xi, \eta)$ to the discrete indices $(m,n)$. The detector integrates the intensity over the whole pixel whereas the model just samples discretely at each pixel. With greater than Nyquist discretization ($\geq$ 2 pixels per $\lambda$/D) of the PSF already required for wavefront correction, the effect of sampling the PSF at each pixel instead of integrating over that area is small. The discretized focal plane electric field is thus
\begin{align}
E_{k,m,n}  &= E_{k-1,m,n} + \sum_q^{N_{act}}  \Delta u_{k,q} B_{k-1,q,m,n},
\label{eq:EfocDMTaylorDiscrete}          
\end{align}
where we have implicitly defined the region as being only within the dark hole. To perform matrix operations on the discretized field, it is convenient to reshape the field into a vector of length $N_{pix}$, the number of dark hole pixels, such that
\begin{align}
E_{k}  &= E_{k-1} + G_{k-1} \Delta u_{k}.
\label{eq:EfocDMvector}          
\end{align}
 Both $E_{k}$ and $E_{k-1} $ have dimensions $N_{pix} \times 1$, the control Jacobian $G_{k-1}$ has dimension $N_{pix} \times N_{act}$, and the vector of control commands $\Delta u_{k}$ has dimension $N_{act} \times 1$.

Setting Eq.\ \ref{eq:EfocDMvector} equal to zero and solving for $\Delta u_{k}$, we obtain the command to minimize the dark hole electric field,
 \begin{align}
	 \Delta u_{k} &=-\mathcal{R} \{ G_{k-1}^L E_{k-1}\},
\label{eq:u_opt_energyMin}
\end{align}
where the superscript $L$ gives the left pseudoinverse and $\mathcal{R} \lbrace \cdot \rbrace$ returns the real part. In both EFC and stroke minimization the actuator command is damped to avoid singularities and becomes
 \begin{align}
	 \Delta u_{k} &=-(G_{k-1}^*G_{k-1}+\alpha \mathbb{I})^{-1}\mathcal{R} \lbrace G_{k-1}^*E_{k-1}  \rbrace,
\label{eq:u_opt_EFC}
\end{align}
where $^*$ gives the conjugate transpose, $\alpha$ is the damping (i.e., regularization) value, and $\mathbb{I}$ is the identity matrix. The control Jacobian $G_{k-1}$ is usually not updated ($G_0$ is used instead) to save computation time at the expense of somewhat slower correction. From here on we will use the notation $G$ instead of $G_0$ and $u_k$ instead of $\Delta u_k$ for convenience. The errors in the estimate and model, ignored nonlinearities of the DM phase contribution, and the use of regularization cause the new, corrected field to have only a slight improvement in contrast after each control step. The correction is thus iterative, with a new DM command calculated and applied after each new estimate of the electric field.  

\subsection{Batch Process Pair-wise Estimation} \label{sec:batch}	
The model-based control techniques in the previous section require knowledge of the electric field in the dark hole.  An estimation approach is thus needed to determine the field from intensity measurements in the science camera. Currently the baseline estimation method for a coronagraphic space mission is pair-wise difference imaging as developed by Give'on \cite{giveon2007broadband}, which probes the image via small DM perturbations. It is the only model-based estimation scheme that has attained better than $10^{-8}$ contrast in laboratory experiments,\cite{belikov2007broadband,moody2008design,kern2013piaa, serabyn2013results, riggs2013demonstration} all of which have been in the High Contrast Imaging Testbed (HCIT) at the Jet Propulsion Laboratory (JPL). Pair-wise estimation is characterized by two notable features: it can be used with any coronagraph and it is fairly robust to model uncertainty. This method is described in detail in several papers,\cite{giveon2007broadband, giveon2011pair,groff2015methods} but we revisit the derivation to provide the mathematical foundation for our new work. 

In this paper, we focus on monochromatic wavefront estimation. At both Princeton's HCIL and JPL's HCIT, broadband wavefront correction is accomplished by taking images at several smaller bandpasses within the whole bandpass, creating separate electric field estimates for each one, and then weighting the bandpasses equally within the controller.  

In pair-wise estimation, shapes are actuated on the DM to probe the electric field in the dark hole. Give'on et al.\ \cite{giveon2011pair} explain one such method for choosing probe sets to modulate sufficiently the real and imaginary parts of the field. A separate image is taken for the positive and negative of each probe shape applied on the DM. Let $u_i$ be the differential control signal for the $i$-th positive probe shape. Then, the change in the focal plane electric field from a positive or negative probe is defined as $p_{i\pm}=\pm Gu_i$. For convenience we will not explicitly write the dark hole pixel index for the probe field $p$, the electric field $E$, and the intensity $I$. The focal plane intensity at each dark hole for a given positive or negative probe shape is then
\begin{align}
I_{k,i\pm} &= |E_{k} + p_{k,i\pm}|^2  + n_{k,i\pm} \nonumber \\
&= |E_k|^2 + |p_{k,i}|^2 \pm 2\mathcal{R}\lbrace E_k^*p_{k,i}  \rbrace + n_{k,i\pm},
\end{align}
where $n_{k,i\pm}$ is the zero-mean, Gaussian measurement noise. The difference of the  positive and negative probed images is equal to twice the cross term,
\begin{equation}
\Delta I_{k,i} = I_{k,i+} - I_{k,i-}  =4\mathcal{R}\lbrace E_k^*p_{k,i} \rbrace + n_{k,i},
\end{equation}
where $n_{k,i} = n_{k,i+} - n_{k,i-}$ is the total noise having twice the variance of a single probed image. For a set of measurements from $N_{pp}$ probe pairs, the measurement equation is
\begin{equation}
 \begin{bmatrix} \Delta I_{k,1}  \\ \vdots \\  \Delta I_{k,N_{pp}}  \end{bmatrix}
= 4 \begin{bmatrix}  \mathcal{R} \lbrace  p_{k,1} \rbrace   & \mathcal{I} \lbrace p_{k,1} \rbrace  \\ \vdots & \vdots  \\ \mathcal{R} \lbrace  p_{k,N_{pp}} \rbrace   & \mathcal{I} \lbrace  p_{k,N_{pp}} \rbrace  \end {bmatrix}  
    \begin{bmatrix}  \mathcal{R} \lbrace E_k  \rbrace   \\  \mathcal{I} \lbrace E_k  \rbrace  \end{bmatrix}
     +  \begin{bmatrix} n_{k,1}  \\ \vdots \\  n_{k,N_{pp}}  \end{bmatrix},
\label{eq:measEqBatch}
\end{equation}
where $\mathcal{I} \lbrace \cdot \rbrace$ takes the imaginary part of the complex value. We re-write Eq.\ \ref{eq:measEqBatch}  as
\begin{equation}
z_k = H_k x_k + n_k,
\end{equation}
where the set of measurements is
\begin{equation}
 z_k = \begin{bmatrix} \Delta I_{k,1}  \\ \vdots \\  \Delta I_{k,N_{pp}}  \end{bmatrix},
\label{eq:zBatch}
\end{equation}
the linear observation matrix is 
\begin{equation}
H_k = 4 \begin{bmatrix}  \mathcal{R} \lbrace  p_{k,1} \rbrace   & \mathcal{I} \lbrace p_{k,1} \rbrace  \\ \vdots & \vdots  \\ \mathcal{R} \lbrace  p_{k,N_{pp}} \rbrace   & \mathcal{I} \lbrace  p_{k,N_{pp}} \rbrace \end {bmatrix},
\label{eq:HBatch}
\end{equation} 
the state vector is 
\begin{equation}
x_k =  \begin{bmatrix}  \mathcal{R} \lbrace E_k  \rbrace   \\  \mathcal{I} \lbrace E_k \rbrace  \end{bmatrix},
\label{eq:xBatch}
\end{equation}
and the measurement noise vector is
\begin{equation}
n_k =   \begin{bmatrix} n_{k,1}  \\ \vdots \\  n_{k,N_{pp}}  \end{bmatrix}.
\label{eq:BPnoiseVec}
\end{equation}

The best estimate $\hat{x}_k$ of the field's real and imaginary parts is found by taking the left pseudo-inverse of $H_k$,
\begin{equation}
\hat{x}_k = H_k^L z_k,
\label{eq:batchEst}
\end{equation}
which requires two probe pairs to be invertible and at least three probe pairs for a least-squares estimate to reduce error from measurement noise.

In addition to the pairs of probed images, we always take an unprobed image $I_{k}$ to measure the current contrast level. The more critical role of the unprobed image is in the empirically-based estimate of the probe amplitude, 
\begin{equation}
\widehat{|p_{k,i}|} = \sqrt{ \frac{I_{k,i+} + I_{k,i-}}{2} - I_{k}  }.
\label{eq:probeAmp}
\end{equation}
As described by Give'on et al.,\cite{giveon2011pair} this technique mitigates several types of model error to enable faster, deeper correction. Even with a good model of the laboratory, the measured and modeled probe amplitudes can have different morphologies and differ by several times in magnitude. The phase of the probe is still calculated using the model. 

In this batch process estimation, there is an implicit assumption that the wavefront is static. The estimator and controller can still create a dark hole as long as the electric field is static at the level of the contrast target over the course of a few correction iterations.

\subsection{Recursive Pair-wise Estimation with a Kalman Filter} \label{sec:KF} 	
Groff and Kasdin\cite{groff2013filtering} incorporated pair-wise estimation into a Kalman filter for better accuracy and robustness. The KF optimally utilizes the previous estimate, the expected model uncertainty, the control signal, and new measurements in the estimate calculation. Since the KF knows the previous estimate, it does not require a full, invertible set of new measurements. Therefore, the wavefront estimate can be updated with just three new images: one unprobed image and one pair of probed images. If the estimate can be as accurate with fewer images of the same exposure time, this technique can further increase the speed of wavefront correction.\cite{groff2013filtering, riggs2013demonstration,riggs2015afta} The KF formulation also permits a dynamic state, so the KF can correct a dynamic wavefront faster and more robustly than a batch process estimator.    

 \subsection{Batch Estimation of Incoherent Light} \label{sec:batchIincoest} 

Both formulations of pair-wise estimation (batch and recursive) can yield a batch estimate of the incoherent light intensity at each correction iteration $k$. The incoherent intensity estimate $\hat{I}_{inco,k}$ at each pixel is 
 \begin{equation}
 \hat{I}_{inco,k} = I_{k}-|\hat{E}_{k}|^2,
 \label{eq:batchIinco}
\end{equation}
where $\hat{E}_{k}$ is the estimated stellar electric field. We will not derive the variance for the starlight intensity here, but from Eq.\ \ref{eq:batchIinco} we can see that the incoherent intensity batch estimate has a higher variance than a single image. Both terms in Eq.\ \ref{eq:batchIinco} are susceptible to noise sources (shot, readout, and dark current noise), and the estimated starlight intensity is susceptible to model errors. To mitigate both model errors and measurement noise, it would be better to estimate the incoherent light recursively with a filter that can use previous data and appropriate weights on the error sources.

\section{Recursive Estimation of Both Incoherent Light and the Stellar Wavefront} \label{sec:recursive}

The batch and recursive pair-wise estimators in Sections \ref{sec:batch} and \ref{sec:KF} have been tested and proven to work at high contrast for suppressing coherent, on-axis light in JPL's HCIT.\cite{giveon2011pair, riggs2013demonstration} The ultimate goal of wavefront correction, however, is to image faint sources that are incoherent with a star such as exoplanets and disks. During wavefront correction the starlight speckles change as they are suppressed while the exoplanets and disks remain unchanged, so a recursive filter can form a better estimate of the incoherent light with each new set of images. By implementing Bayesian techniques to locate any exoplanets or disks in the incoherent image,\cite{kasdin2006bayesian} we can better detect and characterize our science targets with just the correction images.

One possible approach to recursive incoherent estimation is to use another KF on the batch incoherent estimates from Eq.\ \ref{eq:batchIinco}. While this method would let us use two linear estimators, it is inefficient because the estimates of the incoherent light and starlight are interdependent. To produce the best estimate of the incoherent light with all available data, we need to estimate the stellar electric field and incoherent intensity simultaneously in a nonlinear estimator.

With a nonlinear filter, one could attempt to utilize the true nonlinear phase dependence of the electric field on the DM surface,
\begin{align}
E_k(\xi,\eta) &= \mathcal{C}\lbrace \tilde{E}_{0}(x,y) e^{i \phi_{k-1}(x,y)} e^{i \Delta \phi_k(x,y)} \rbrace,
\label{eq:gofu}          
\end{align}
in the modeled propagation of the electric field, but we do not. Because opaque coronagraphic masks and field stops between the DM and camera block light, we cannot directly back-propagate our electric field estimate from the focal plane to the DM-plane. Maximum a posteriori methods such as COFFEE exist to solve this problem, but they are currently too slow computationally to implement in real time. The other reason for not utilizing the full nonlinear model and for not including more terms in the Taylor expansion in Eq.\ \ref{eq:EpupDMTaylor} is that the errors in our knowledge of $\mathcal{C}\lbrace \cdot \rbrace$, $\tilde{E}_{0}(x,y)$, $\phi_{k-1}(x,y)$, and $\Delta \phi_k(x,y)$ might outweigh the better accuracy from a higher-order model. 

As a first step into nonlinear focal plane wavefront estimators, we derive an extended Kalman filter (EKF) as first shown by Riggs et al.\cite{riggs2014estimation} In this paper, the EKF utilizes the same probing strategy as the KF does for easier performance comparison. The EKF has the advantage that it can utilize the unprobed image recursively as well. 

 
\subsection{Constructing the EKF}

 We augment the original state vector in Eq.\ \ref{eq:xBatch} to include  the incoherent intensity at each pixel, $I_{inco, k}$,
 \begin{align}
	x_{k} &= \begin{bmatrix} \R \{ E_{k} \} \\  \I \lbrace E_{k} \rbrace \\ I_{inco, k} \end{bmatrix}. 
\label{eq:xEKF}
\end{align} 
 
The most general EKF measurement vector is the actual set of images taken. This formulation allows the use of unpaired probes or multi-DM probes, which we leave for future work. Here we use the same set of images as in pair-wise estimation such that $z$ at each dark hole pixel consists of the unprobed image $I_{k}$ and the $2{\times} N_{pp}$ probe images, 
 \begin{align}
	z_{k} &= \begin{bmatrix} I_{k} \\  I_{k,1+} \\ I_{k,1-} \\ \vdots \\   I_{k,N_{pp}+} \\ I_{k,N_{pp}-} \end{bmatrix} \nonumber \\
	    &= h(x_{k}) +  n_{k} + \mathcal{O}\lbrace \Delta \phi_k^2 \rbrace,
\label{eq:ndEKFz}
\end{align} 
where $h(x_{k})$ is the nonlinear measurement function and $\mathcal{O}\lbrace\Delta \phi_k^2 \rbrace$ is the model error from ignored higher-order terms of the DM-phase Taylor series. The additive measurement noise vector $n_{k}$ consists of readout noise, photon shot noise, and dark current. By not performing pair-wise differencing of the probed images, the terms of order $\Delta \phi_k^2$  ignored in Eq.\ \ref{eq:EpupDMTaylor} no longer cancel and appear in the measurement. The quadratic, approximate measurement function is
 \begin{align}
 h(x_{k})  &=  \begin{bmatrix} | E_{k} |^2 + I_{inco, k}  \\
 			| E_{k}  + G u_1 |^2 + I_{inco, k}  \\
			| E_{k}  - G u_1 |^2 + I_{inco, k}  \\
			\vdots \\
			| E_{k}  + G u_{N_{pp}} |^2 + I_{inco, k}  \\
	    		| E_{k}  - G u_{N_{pp}} |^2 + I_{inco, k}  		
			\end{bmatrix} \nonumber \\
 &=  \begin{bmatrix} (\R \{ E_{k} \})^2 + (\I \{ E_{k} \})^2 + I_{inco, k}  \\
 			(\R \{ E_{k}+ G u_1 \} )^2 + (\I \{ E_{k}+ G u_1 \})^2 + I_{inco, k}  \\
			(\R \{ E_{k}- G u_1 \} )^2 +  (\I \{ E_{k}- G u_1 \})^2 + I_{inco, k}  \\
			\vdots \\
 			(\R \{ E_{k}+ G u_{N_{pp}}  \} )^2 + (\I \{ E_{k}+ G u_{N_{pp}}  \})^2 + I_{inco, k}  \\
			(\R \{ E_{k}- G u_{N_{pp}}  \} )^2 +  (\I \{ E_{k}- G u_{N_{pp}}  \})^2 + I_{inco, k}  \\		
			\end{bmatrix} \nonumber \\
 &=  \begin{bmatrix} (x_{k}[1])^2 + (x_{k}[2])^2 + x_{k}[3]  \\
	    		(x_{k}[1] + \mathcal{R} \lbrace G \rbrace u_1)^2 + (x_{k}[2]+ \mathcal{I} \lbrace G \rbrace u_{1})^2 + x_{k}[3] \\
			(x_{k}[1] - \mathcal{R}\lbrace G \rbrace u_1)^2 + (x_{k}[2]- \mathcal{I}\lbrace G \rbrace u_1)^2 + x_{k}[3] \\
			\vdots \\
	    		(x_{k}[1] + \mathcal{R} \lbrace G \rbrace u_{N_{pp}})^2 + (x_{k}[2]+ \mathcal{I} \lbrace G \rbrace u_{N_{pp}})^2 + x_{k}[3] \\
			(x_{k}[1] - \mathcal{R}\lbrace G \rbrace u_{N_{pp}})^2 + (x_{k}[2]- \mathcal{I}\lbrace G \rbrace u_{N_{pp}})^2 + x_{k}[3] 			
			\end{bmatrix},
\label{eq:ndEKFhofx}
\end{align} 
where $x_k[m]$ represents the $m$-th element of vector $x_k$, and $u_i$ is the additive DM control signal for the positive probe shape $i$.

To derive the EKF, we first re-define the true dynamic state equation at each image plane pixel as
 \begin{align}
	x_k &= \Phi {x}_{k-1} + \Gamma u_{k-1}  + \Lambda w_{k-1} \label{eq:stateDynamics},
\end{align} 
where $\Phi$ is the state transition matrix, $\Gamma$ is the real-valued control Jacobian, and $\Lambda$ is the disturbance Jacobian. The variable $w_{k-1}$ is random process noise; it is included in the model to accommodate model errors and random, unknown disturbances. We treat our system as static, so $\Phi$ is just the identity matrix. The only source of change is from the DMs, which means the only source of model error is in our knowledge of the DM response. Thus, $\Lambda=\Gamma_{true}-\Gamma_{model}$ and $w_{k-1}=u_{k-1}$. From here on $\Gamma$ will mean $\Gamma_{model}$. The third row of $\Gamma$ is zeroes because the incoherent light is not modulated by the DMs. (Only the PSF core is observable for faint incoherent sources, and high-order wavefront correction primarily changes the wings of the PSF.)

The EKF minimizes both the error of the state estimate and the state covariance estimate. The state covariance $P$ is defined as the expectation value $E[\cdot]$ of the outer product of the error in the state estimate,
 \begin{align}
P_{k}=E[(x_{k}-\hat{x}_{k})(x_{k}-\hat{x}_{k})^T].
\end{align}

In the first two equations of the EKF, the dynamics of the system are used to propagate the previous estimates of the state and state covariance to the current time step. Following the derivation and notation by Stengel,\cite{stengel1994optimal} the state estimate time update is
 \begin{align}
	\hat{x}_k(-) &= \Phi \hat{x}_{k-1}(+) + \Gamma u_{k-1},
	\label{xextrap}
\end{align} 
and the covariance estimate time update is
 \begin{align}
	P_k(-) &= \Phi P_{k-1}(+)\Phi^T + Q_{k-1}, 
	\label{Pextrap}
\end{align} 
where $Q_{k-1}$ is the process noise matrix. The signifier $(-)$ means $\hat{x}_k$ or $P_k$ is the model-based time update, and the signifier $(+)$ means it is the measurement-updated  estimate for that correction iteration. We assume that the unknown disturbance $\Lambda w_{k-1}$ is Gaussian and zero mean so it should not change the expected value $\hat{x}_k$. The process noise covariance matrix is given by
 \begin{align}
	Q_{k-1}=\Lambda E[w_{k-1} w_{k-1}^T] \Lambda ^T. \label{Qdef}
\end{align} 

The last stage of the EKF is to improve the estimates with new data in the measurement update equations, 
 \begin{align}
	\hat{x}_k(+) &= \hat{x}_k(-) + K_k[z_k - h(\hat{x}_k(-))] \label{xUpdate} \\
	P_k(+) &= [I-K_k H_k]P_k(-), \label{Pupdate}
\end{align}
where the Kalman gain $K_k$ optimally balances the weighting of the model error and old data versus the new measurements. The Kalman gain is defined as
 \begin{align}
	K_k &= P_k(-) H_k^T[H_k P_k(-) H_k^T + R_k]^{-1} \label {KEKF},
\end{align} 
where $R_k$ is the measurement noise covariance matrix, which we discuss in more detail in Section \ref{QandR}. $H_k$ is the observation matrix linearized about the state time update,
 \begin{align}
	H_k &= \frac{\partial h(x)}{\partial x} \biggr\vert_{x=\hat{x}_k(-)} \nonumber \\
	&=  \begin{bmatrix}  2x[1] & 2x[2] & 1 \\
	  2(x_{}[1] + \mathcal{R} \lbrace G \rbrace u_1) & 2(x_{}[2] + \mathcal{I} \lbrace G \rbrace u_1) & 1 \\
	  \vdots & \vdots & \vdots \\
	 2(x_{}[1] - \mathcal{R} \lbrace G \rbrace u_{N_{pp}}) & 2(x_{}[2] - \mathcal{I} \lbrace G \rbrace u_{N_{pp}}) & 1 \end{bmatrix} \Biggl\lvert_{x=\hat{x}_k(-)}.
\label{eq:ndobsv1pair}
\end{align}

In summary, the five EKF equations for our formulation are
 \begin{align}
	\hat{x}_k(-) &= \hat{x}_{k-1}(+) + \Gamma u_{k-1} \label{xextrap2} \\
	P_k(-) &= P_{k-1}(+) + Q_{k-1} \\
	K_k &= P_k(-) H_k^T[H_k P_k(-) H_k^T + R_k]^{-1} \\
	\hat{x}_k(+) &= \hat{x}_k(-) + K_k[z_k - h(\hat{x}_k(-))] \\
	P_k(+) &= [\mathbb{I}-K_k H_k]P_k(-). \label{Pupdate2}
\end{align} 
We summarize the variables used in these equations in Table \ref{EKF1pairvardim}, and we list the matrices and their definitions in Table \ref{EKF1pairmatrixdim}. The estimate is performed separately at each dark hole pixel to avoid the use of extremely large matrices.

 \begin{table}[h]
\begin{center}
\renewcommand{\arraystretch}{1.5}
\begin{tabular}{| C{5cm}  c  c |} 
\hline
 Variable & Representation & Dimension \\
 \hline
State Estimate & $\hat{x}_{k} = \begin{bmatrix} \mathcal{R} \lbrace E_{k}   \rbrace \\  \mathcal{I} \lbrace E_{k} \rbrace \\ I_{inco, k} \end{bmatrix} $ & $3 \times 1$ \\
Intensity Measurements & $z_{k}$ & $N_z \times 1$ \\
Sensor Noise & $n_{k}$ & $N_z \times 1$ \\
DM Commands & $u_k$ & $(N_{DMs}\times N_{act})\times 1$ \\
Process Noise & $w_k$ & $(N_{DMs}\times N_{act})\times 1$  \\
\hline
\end{tabular}
\setlength{\tabcolsep}{10pt}
\end{center}
\caption{Dimensions of variables for the EKF. $N_z = 1+2N_{pp}$. }
\label{EKF1pairvardim}
\end{table}

 \begin{table}[h]
\begin{center}
\renewcommand{\arraystretch}{1.5}
\begin{tabular}{| C{5cm}  c  c |} 
\hline
 Matrix & Representation & Dimension \\
 \hline
Linearized State Response & $\Phi=\mathbb{I}$ & $3 \times 3$ \\
Nonlinear Observation & $h(x)$ & $N_z \times 1$ \\
Linearized Observation & $H_{k}=  \frac{\partial h(x)}{\partial x} \bigr\vert_{x=\hat{x}_k(-)}$ & $N_z \times 3$\\
Linearized Complex Response of Probing DM & $G$ & $1 \times N_{act}$ \\
Linearized Response of Probing DM & $\Gamma= \begin{bmatrix}  Re\lbrace G[1] \rbrace \dotsb Re\lbrace G[N_{act}] \rbrace  \\
	Im\lbrace G[1] \rbrace \dotsb Im\lbrace G[N_{act}] \rbrace \\
	0 \dotsb 0  \end{bmatrix}$ & $3 \times N_{act}$ \\
Disturbance Response & $\Lambda=\Gamma$ &  $3 \times N_{act}$  \\	
State Covariance (Time Update) & $P_{k}(-)=E[(x_{k}-\hat{x}_{k}(-))(x_{k}-\hat{x}_{k}(-))^T]$ & $3\times 3$ \\
State Covariance (Measurement Update) & $P_{k}(+)=E[(x_{k}-\hat{x}_{k}(+))(x_{k}-\hat{x}_{k}(+))^T]$ & $3\times 3$ \\
Process Noise & $Q_k=\Lambda E[w_k w_k^T] \Lambda ^T$ & $3 \times 3$ \\
Sensor Noise & $R_{k}=E[n_{k} n_{k}^T]$ & $N_z \times N_z$ \\
Kalman Gain & $K_{k}$ is computed & $3 \times N_z$ \\
\hline
\end{tabular}
\setlength{\tabcolsep}{10pt}
\end{center}
\caption{Dimensions of matrices for the EKF. }
\label{EKF1pairmatrixdim}
\end{table}

\subsection{Sensor and Process Noise} \label{QandR} 	

In any Kalman filter, $R_k$ and $Q_{k-1}$ are the tuning parameters. The sensor noise matrix is defined as the expectation value of the outer product of the measurement noise vector,
 \begin{align}
	R_{k}= E[n_{k} n_{k}^T].
\end{align} 
 We can calculate the value of $R_k$ based on camera measurements, so only $Q_k$ needs to be tuned. The main sources of measurement noise are dark current, readout noise, and photon shot noise. The total variance in Analog-Digital Units (ADU, or counts) expected at each pixel is
 \begin{align}
	\sigma_{total}^2 = (c_k f_{star} t_{exp} + \sigma^2_{ron} + (s_{dark} t_{exp}) )/(g n_{exp}),
\label{eq:Rstd}
\end{align} 
where $g$ is the gain of the detector in photoelectrons/ADU, $c_k$ is the average measured contrast in the dark hole, $f_{star}$ is the peak flux of the starlight in photoelectrons/second, $t_{exp}$ is the exposure time per frame, $\sigma_{ron}^2$ is the variance of the readout noise in photoelectrons, $s_{dark}$ is the dark current rate in photoelectrons/second, and $n_{exp}$ is the number of exposures averaged to make an image. The contrast across the dark hole in either the probed or unprobed images are relatively uniform, so we use the same matrix $R_k$ at each pixel. We still use separate values for probed or unprobed images since the probed images have more light. The noise from image to image is uncorrelated, so $R_k$ is a diagonal matrix. This means that each diagonal entry $r_k$ in $R_k$ is simply the variance in units of contrast,
 \begin{align}
	r_k = \sigma^2_{k, total}/(f_{star} t_{exp}).
\label{eq:Rvariance}
\end{align} 
The sensor noise matrix is then
 \begin{align}
	R_{k}
	=  \begin{bmatrix}  	r_{k,unpr}  &  & & 0  \\
						  & r_{k,pr}  &  &  \\
						 &    & \ddots &   \\
	  					0 &  &  & r_{k,pr} \\
	   \end{bmatrix},
\label{eq:Rdef}
\end{align} 
where $r_{k,unpr}$ is for the unprobed image and $r_{k,pr}$ is for the probed images.



We must include a nonzero process noise $Q_{k-1}$ in the covariance estimate extrapolation because of the uncertainty in the control step. Although we assume that the DM does not modulate the incoherent light, we must still include process noise to prevent the filter from converging quickly to an incorrect value. We scale the process noise for the third state with the average incoherent intensity estimate. Without location-specific information of the process noise, we assign the same $Q_{k-1}$ matrix to each image plane pixel. Similarly, we have no way of distinguishing if the real or imaginary parts of the starlight should have more or less model error, so we set those covariance values as equal. For each pixel at correction iteration $k$, we thus use the process noise matrix 
 \begin{align}
	Q_{k-1}  =  \begin{bmatrix}  	q_0 |\hat{E}_{k-1}|^2_{avg}  & 0 & 0 \\
						0 & q_0 |\hat{E}_{k-1}|^2_{avg}  & 0 \\
	  					0 & 0 & q_{3} (\hat{I}_{inco,k-1})^2_{avg} \\
	   \end{bmatrix},
\label{eq:Qdef}
\end{align} 
where $q_0$ and $q_3$ are constants used to tune the relative values. There should also be off-diagonal elements in $Q_k$ since the model errors of the states are cross-correlated and there is unmodeled inter-actuator coupling, but we nevertheless keep $Q_{k-1}$ diagonal to avoid dropping rank before the inversion in the Kalman gain calculation. With good tuning, the values of $P_{k}(+)$ tend to show no cross-correlation (zero values) among the electric field and incoherent states but a slight cross-correlation (about 10\% of the diagonal values) between the real and imaginary parts of the field. Including this nonzero off-diagonal term in $Q_{k-1}$ did not change performance of the EKF, and was therefore not included for all the tests reported in Section \ref{sec:Results}.

\subsection{Iterated Extended Kalman Filter} 	

When using pair-wise differencing for the starlight measurements, the linear observation matrix $H_k$ is correct to third order in the DM-phase Taylor series expansion. In our EKF formulation without differencing, the quadratic terms no longer cancel. The linearization of the observation $h(x)$ in Eq.\ \ref{eq:ndobsv1pair} thus depends on the current state, so it is necessary to use the initial, inaccurate time update $\hat{x}_{k}(-)$ as the linearization point.

A large body of research already exists to address nonlinear errors when implementing an EKF. Here we try the simplest improvement, which is to iterate the EKF (known as an IEKF) to mitigate nonlinearities. The main error in $H_k$ (and subsequently in $K_k$, $\hat{x}_{k}(+)$, and $P_{k}(+)$ ) comes from the linearization about the model-based time update $\hat{x}_{k}(-)$, but after running the EKF (Eqs.\ \ref{xextrap2}-\ref{Pupdate2}) we have a more accurate estimate of the state available. Using $\hat{x}_{k}(+)$ as the new linearization point for an updated $H_k$, the IEKF recomputes $K_k$, $\hat{x}_{k}(+)$, and $P_{k}(+)$. There is now an even better estimate of the state, and this process of iterating the EKF can be repeated until the state estimate converges on a solution. Defining the subscripts for the EKF iterations as $j=0,1,2,...,N_{it}$, we follow the notation of Gelb\cite{gelb1974optimal} and Simon\cite{simon2006optimal} and write the IEKF equations as
 \begin{align}
 	H_{k,j} &= \frac{\partial h(x)}{\partial x} \biggr\vert_{x=\hat{x}_{k,j}(+)}   \label{HupdateIEKF} \\
	K_{k,j} &= P_k(-) H_{k,j}^T[H_{k,j} P_k(-) H_{k,j}^T + R_k]^{-1} \\
	\hat{x}_ {k,j+1}(+) &= \hat{x}_k(-) + K_{k,j}\Bigl(z_k - h(\hat{x}_{k,j}(+)) -H_{k,j}[\hat{x}_ {k}(-)-\hat{x}_ {k,j}(+)] \Bigr) \label{eq:IEKFxMeasUpdate}\\
	P_{k,j+1}(+) &= [\mathbb{I}-K_{k,j} H_{k,j}]P_k(-). \label{PupdateIEKF}
\end{align} 
We initialize the IEKF with
 \begin{align}
	\hat{x}_ {k,j=0}(+) &= \hat{x}_k(-) \\
	P_{k,j=0}(+) &= P_k(-)
\end{align}
and then iterate the filter by updating Eqs.\ \ref{HupdateIEKF}-\ref{PupdateIEKF} to converge on a better state estimate at the $k$\textsuperscript{th} control step. If $N_{it}=0$, the IEKF simplifies back to the EKF. In Section \ref{sec:accuracy} we test several values of $N_{it}$ and determine the minimum value to converge on an estimate.

\section{Computational Complexity} \label{sec:Complexity}
 
 Space-based observatories have limited computing power, so it is important to compare the computational complexity of the estimators proposed. In pair-wise estimation the state can be estimated separately at each image-plane pixel, which means that only small matrices are required but the calculations must be repeated thousands of times. Here we derive estimates of the complexity for each estimator based on the number of matrix multiplications (and divisions) required per pixel. Because there are many possible methodological (such as taking another image during calculations) or algorithmic (such as using alternate forms of equations) approaches to reduce the effective complexity of the estimators, we perform just a simple analysis as a starting point for comparison. We leave out the re-calculation of the observation matrix, $H_k$, for each estimator because calculating the new values of $\mathcal{R} \lbrace G \rbrace u_{k,i}$ and $\mathcal{I} \lbrace G \rbrace u_{k,i}$ is common to all the estimators and requires the square root calculation in Eq.\ \ref{eq:probeAmp}. It should also be noted that the recursive estimators require more memory, but that is a separate issue from the processing speed considered here.

Table \ref{tab:complexity} shows the relative complexity of each estimator in terms of floating point multiplications required per pixel. The batch process calculation is based on Eq.\ \ref{eq:batchEst} assuming that the minimum of two probe pairs is used. The total number of multiplications is 26. 

 \begin{table}[h]
\begin{center}
\renewcommand{\arraystretch}{1.5}
\begin{tabular}{| c |c |} 
\hline
 Estimator & \# of Multiplications  \\
 \hline
BP (2p) & 26 \\
\hline
KF (1p) & 19 \\
KF (2p)     &  46 \\
\hline
EKF (1p) & 150 \\
EKF (2p) & 360 \\      
\hline
IEKF (1p) & $150+159 N_{it}$ \\
IEKF (2p) & $360+375 N_{it}$ \\           
\hline
\end{tabular}
\setlength{\tabcolsep}{10pt}
\end{center}
\caption{Number of scalar multiplications (and divisions) required per image-plane pixel for each estimator. The number of probe pairs used are shown in parentheses. BP stands for batch process, and $N_{it}$ is the number of EKF iterations. }
\label{tab:complexity}
\end{table}

The KF equations have the same form as Eqs.\ \ref{xextrap2}-\ref{Pupdate2} except the state estimate update has a linear observation,
 \begin{align}
	\hat{x}_k(+) &= \hat{x}_k(-) + K_k[z_k - H_k \hat{x}_k(-))]. \label{xUpdateKF} 
\end{align}
The KF has only two states, so the dimensions of the KF matrices are the same as for the EKF in Table \ref{EKF1pairmatrixdim} except each 3 should be a 2. The most computationally expensive calculation in the KF is the multiplication of $ \Gamma u_{k-1}$ in the time update of the state estimate, which requires $2 N_{act}$ multiplications. For WFIRST-AFTA, this would be about three thousand. The controller should already have performed this calculation to choose the optimal DM command, so we assume that it adds no complexity to the estimator. We find that the 1-pair KF requires 19 multiplications, so it is slightly less computationally expensive than the batch process. The 2-pair KF, requiring 46 multiplications, needs fewer than twice the number of computations in the 2-pair batch process.

The EKF as listed in Eqs.\ \ref{xextrap2}-\ref{Pupdate2} requires many more calculations than the KF because of the longer state and measurement vectors. We find that the 1-pair EKF requires 150 multiplications and the 2-pair EKF requires 360. These numbers could be reduced by exploiting the sparsity of the matrices and not performing a brute force matrix inversion in the Kalman gain calculation. Iterating the filter requires several more calculations per EKF iteration because of the extra multiplication in Eq.\ \ref{eq:IEKFxMeasUpdate}. The IEKF requires on the order of a thousand multiplications at each of the few thousand pixels. Since the inversion in Eq.\ \ref{eq:u_opt_EFC} of the controller can be precomputed, the IEKF and controller would each require on the order of a million calculations per correction iteration. We conclude that the IEKF should still be feasible for the limited computing power available on a space observatory.   
\section{Experimental Results} \label{sec:Results}
 
 In this section, we present experimental validation of our EKF and IEKF formulations in the High Contrast Imaging Laboratory (HCIL) at Princeton. We use the stroke minimization controller with fixed settings, so that any variation in performance should arise from estimator. We compare the EKF and IEKF results to those for the batch process estimator and KF with and without incoherent sources present. Finally, we identify the limitations in our lab to be addressed in future work. 
 
\subsection{High Contrast Imaging Laboratory at Princeton} 	

\begin{figure}[ht!] 
\centering
    \includegraphics[width = .7\columnwidth,trim=.05in 0in .05in 0in]{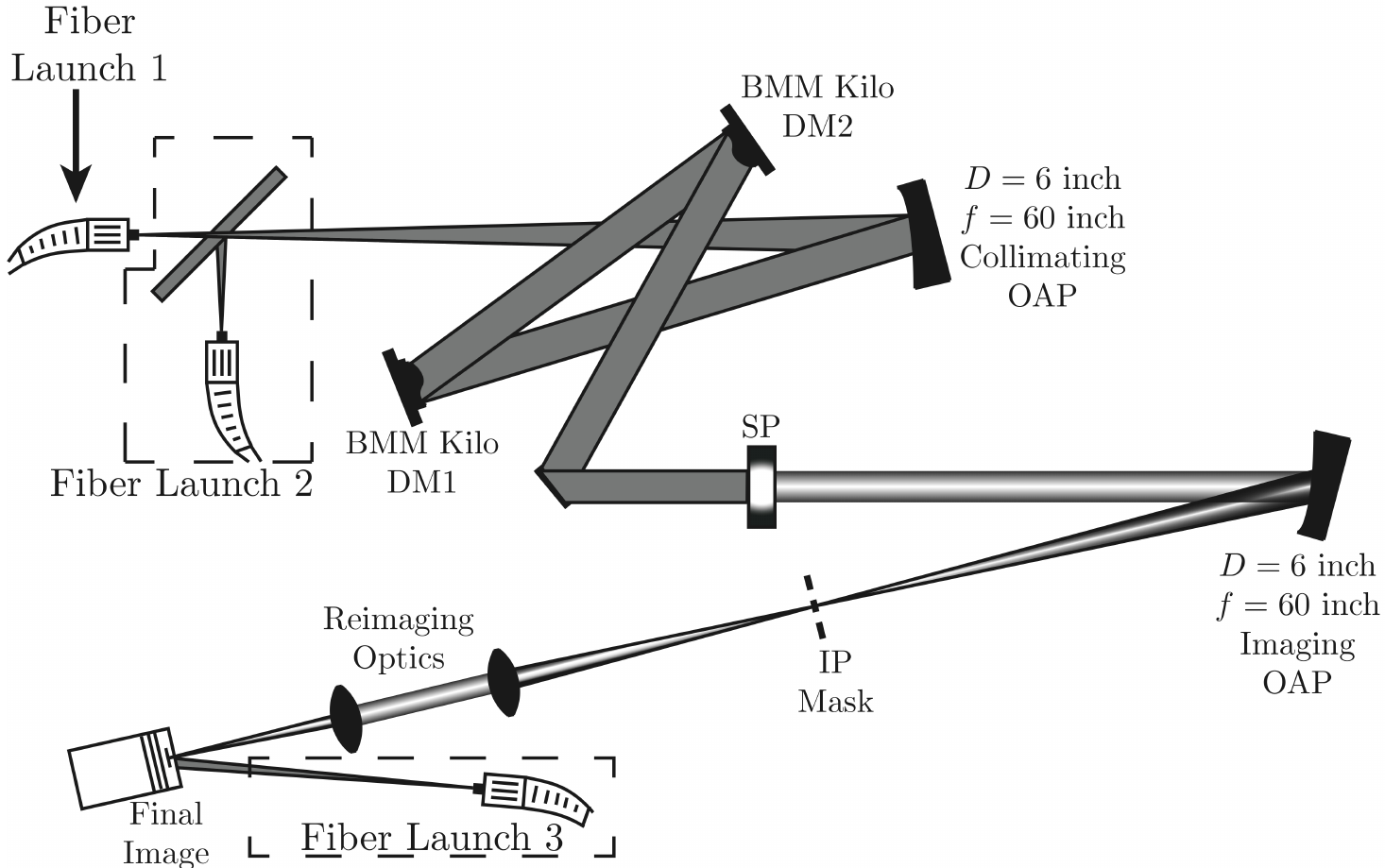}
    \caption{Diagram of the HCIL configuration. Dashed boxes show the modified configurations with additional fiber launches to introduce incoherent sources. The beamsplitter and fiber launch 2 are used to inject an off-axis, planet-like source. Fiber launch 3 adds a zodi-like, flat background in the focal plane.}
     \label{HCILlayout}
\end{figure}

\begin{figure}[ht!]   
\centering
    \includegraphics[width = .8\columnwidth,trim=.05in 0in .05in 0in]{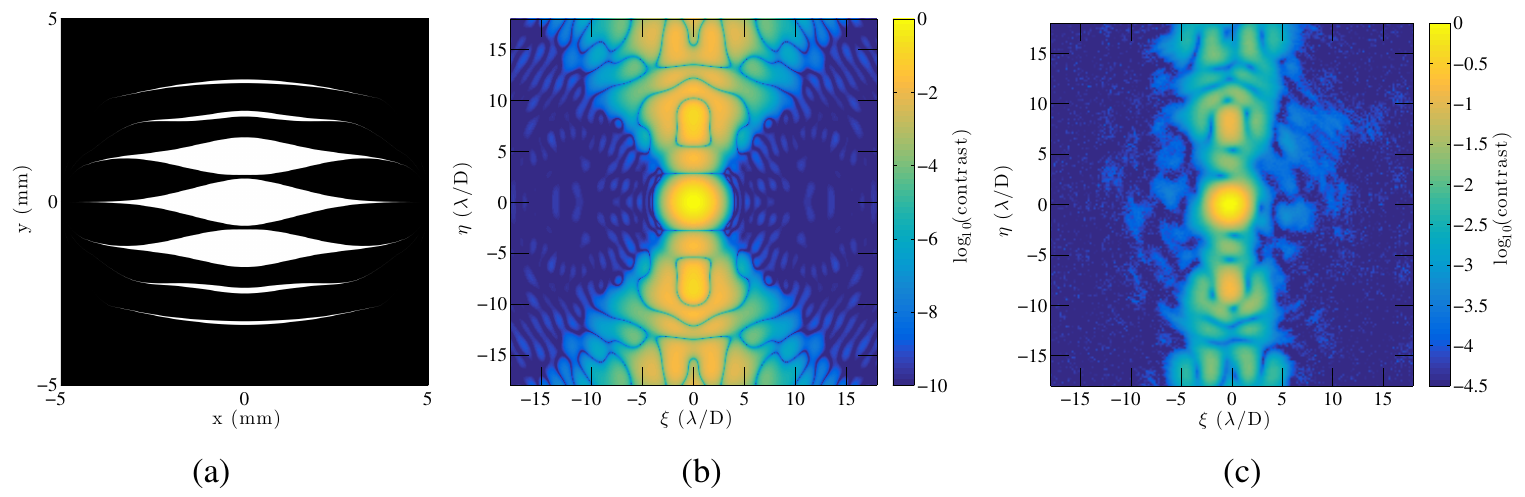}    
\caption{ (a) The Ripple3 shaped pupil used in the HCIL along with (b) its normalized design PSF on a log scale. The ideal average contrast is $3{\times}10^{-10}$ from $4-40 \lambda/D$ over symmetric $90^{\circ}$ sectors. (c) The uncorrected PSF as measured in the HCIL is shown with the same spatial scaling but a shorter log stretch.}
      \label{fig:SPandPSF}
\end{figure}

In the HCIL, we utilize shaped pupil (SP) coronagraphs to generate high contrast. Our layout, as shown in Fig.\ \ref{HCILlayout}, uses as few optics as possible to enable easier alignment and introduce fewer optical aberrations. 
We inject monochromatic, 635-nm laser light directly from a fiber (launch 1) as the simulated stellar point source in the nominal experimental configuration. The 60-inch focal length of the off-axis parabola (OAP) allows us to approximate the central part of the beam as uniform. The collimated beam reflects off two Boston Micromachines Kilo-DMs and a fold mirror in series before reaching a transmissive, 10-mm diameter SP. The SP used in this paper is the freestanding Ripple3 design described by Belikov et al.\cite{belikov2007broadband} and Kasdin et al.\cite{kasdin2007sp} and shown in Fig.\ \ref{fig:SPandPSF}(a). The apodized PSF has a theoretical contrast of $3{\times}10^{-10}$ from $4-40 \lambda/D$ over symmetric $90^{\circ}$ sectors as shown in Fig.\ \ref{fig:SPandPSF}(b), and the empirical, uncorrected PSF in the HCIL is shown in Fig.\ \ref{fig:SPandPSF}(c). The second and final OAP focuses the beam onto a focal plane mask (FPM), which is used only as a field stop for better dynamic range on the detector. Two achromatic lenses then re-image the stopped-down PSF onto a CCD camera.  


To test our estimators in the presence of incoherent sources, we inject additional laser light at either of two locations on our bench as shown in the dashed boxes in Fig.\ \ref{HCILlayout}. To create an exoplanet, we insert a beamsplitter in front of the original fiber source and place fiber launch 2 to reflect into the same beam path. To eliminate any dispersion or path difference errors for the star, launch 1 becomes the planet and launch 2 becomes the star. This is the simplest configuration to add an off-axis source, but the beamsplitter creates additional aberrations and strong polarization dependence. We adjust the planet intensity by using a separate laser source, and we can re-position the planet and star by translating the fiber heads. To approximate a flat zodiacal signal, we place another fiber tip (launch 3) approximately half a meter from the camera. The core of the expanding Gaussian beam is approximately uniform over the detector from this distance. 
       
\begin{figure}[ht!]   
\centering
    \includegraphics[width = .9\columnwidth,trim=.05in 0in .05in 0in]{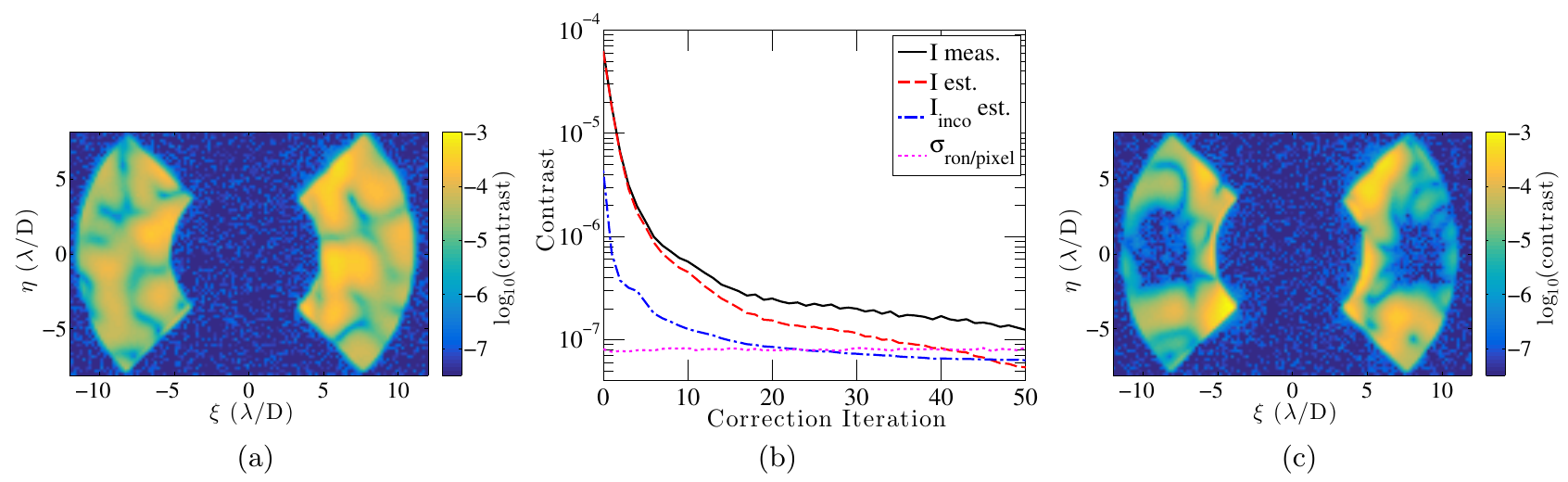}
\caption{An example of a typical correction run, in this case with 2 probe pairs and the IEKF. (a) The stopped-down, uncorrected initial image, shown on a log scale, had a contrast of $6.51 {\times} 10^{-5}$ in the correction region. (c) The final, corrected image had a measured average contrast of $1.2 {\times} 10^{-7}$ in the rectangular dark holes. (b) The contrast correction curve started fast before gradually approaching the highest achievable contrast. The average values are plotted for the measured contrast, estimated starlight contrast, and estimated incoherent contrast in the dark hole, as well as the average standard deviation from readout noise at each pixel, $\sigma_{ron/pixel}$. }
\label{fig:typicalRun}
\end{figure}

We block most of the stellar PSF with a field stop and perform wavefront correction in a subset of the transmitted region. The FPM passes light in symmetric areas from a radius of $5-11 \lambda/D$ over $90^{\circ}$ sectors as shown in Fig.\ \ref{fig:typicalRun}(a). The nominal aberrations set an average starting contrast of $6.51{\times}10^{-5}$ as shown in Fig.\ \ref{fig:typicalRun}(b), in which correction quickly gets the contrast below $10^{-6}$ and then slowly approaches the final achievable contrast of about $10^{-7}$. In these experiments, we corrected only small, rectangular dark holes in the image plane with $\xi \in [-10,-7; 7,10] \ \lambda/D$ and $\eta \in [-2,2] \ \lambda/D$ as shown in the corrected PSF of Fig.\ \ref{fig:typicalRun}(c). When we correct a larger region, we cannot reach as high of a contrast value. 

To compare the relative performance of different estimators, we need to distinguish testbed fluctuations from algorithmic performance. If we perform separate correction runs in our testbed on the same day, we can safely compare them. Otherwise, the optics can drift out of alignment and degrade the correction performance. Because each correction run takes approximately half an hour, we have time for only one or two trials with each estimator when performing comparisons. 

Groff et al.~\cite{groff2015methods} derive the dependence of the electric field estimate variance on the different sources of noise, and we summarize that result here. If the pair-wise probe amplitudes, $|p_{k,i}|$, are much greater than the nominal field amplitude, then the photon shot noise from the starlight can be eliminated. Large probe amplitudes also reduce, but do not eliminate, the variance in the E-field estimate from readout noise and incoherent-light shot noise. If the probe amplitudes are too large, estimate error is introduced from ignored nonlinearities and model error. We manually tuned the probe amplitudes to be as large as possible without slowing correction or limiting the achievable contrast, which gave a probe intensity of about $10^{-6}$ for measured contrast values around $10^{-7}$.  

\subsection{Experimental Goals}	
We sought to determine the accuracy of the EKF and IEKF estimates in several scenarios. Because the true stellar electric field is unobtainable in experiment, we compared the different estimators with contrast correction speed, defined as measured starlight contrast versus total exposure time. We assumed that a better estimate enables a larger correction step. Since the incoherent signal is directly measurable, we quantified the accuracy of the incoherent estimate with the PSF correlation and estimated contrast of an injected exoplanet. Our goals were thus to:
\begin{enumerate}
  \item Compare contrast correction speed and achievable contrast for the IEKF, EKF, KF, and batch process estimator.
  \item Determine the minimum number of EKF iterations needed to get the IEKF state estimate to converge.
  \item Compare the estimators' performance in the presence of zodi-like, incoherent light.
  \item Determine the accuracy of the incoherent estimate by retrieving an injected planet signal.
\end{enumerate} 

\subsection{Experiments without Additional Incoherent Sources}	

\subsubsection{Correction Speed Comparisons}	

\begin{figure}[ht!]  
\centering
    \includegraphics[width = .6\columnwidth,trim=.05in 0in .05in 0in]{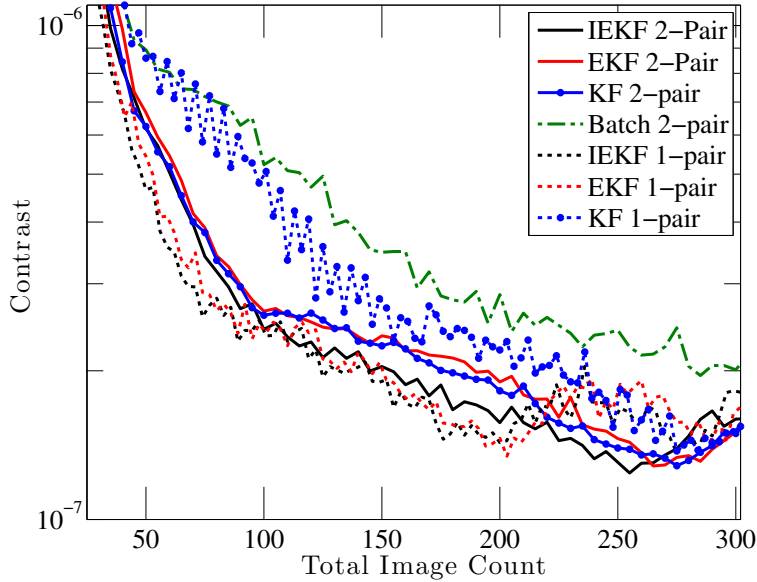}
    \caption{Comparison of contrast correction speed for several estimators without an intentionally injected incoherent source. Exposure time for each image was constant.}
     \label{estCompAll}
\end{figure}

In this experiment, we compared the contrast correction curves for the various estimators as shown in Fig.\ \ref{estCompAll}. The 2-pair estimators used five images per correction iteration (1 unprobed and 2 pairs of probed images) and the 1-pair estimators used three images per correction iteration (1 unprobed and 1 pair of probed images). The same probe shapes were applied every correction iteration for the 2-pair estimators; for the 1-pair estimators the probes alternated after every correction iteration to modulate the field sufficiently. The IEKF performed five EKF iterations.
 
We measured correction speed as the total number of images since all exposures had equal length. The exposure time was the longest possible without saturating any pixels on the detector and gave a contrast resolution of $1.8{\times}10^{-8}$ per count. The batch estimator provided the slowest overall correction. It achieved the worst final contrast and took more correction iterations to reach it. The 1-pair KF was second slowest but did eventually reach as high a contrast as the other recursive estimators. The 2-pair KF and EKF performed equally well, and the 2-pair IEKF performed slightly better than those two after 100 total images. The 1-pair EKF and IEKF started off slightly faster than the others and reached their best achievable contrast in less than 80\% the number of images required for the 2-pair recursive estimators, thereby showing some benefit from fewer probes per correction iteration. Without repeated trials to average out variability, it is important not to draw too many conclusions from these results. In another run (not shown), for example, the 2-pair IEKF performed the same as the 2-pair KF and EKF. Nevertheless, we have observed that the batch process and 1-pair KF are always slower than the other methods and that the 1-pair EKF and IEKF are always slightly faster than all the 2-pair versions. The higher computational complexity of the recursive estimators is partially offset since they require fewer correction iterations to reach a given contrast level.  We have also demonstrated the viability of EKF and IEKF formulations that do not require image differencing.

After reaching the best achievable contrast, each correction curve started to worsen slightly (but did stay at or below about $2{\times}10^{-7}$ contrast). This might have been from the controller being too aggressive or from the modeled control Jacobian not matching the true system well. In a space mission, the system would be better characterized and the controller could be made less aggressive near the ultimate contrast value to prevent divergence.

Although the 1-pair EKF and IEKF slightly outperformed the 2-pair recursive estimators, in later tests we used the 2-pair versions of the estimators. We found in various trials (not shown) that the 2-pair estimates were slightly less sensitive to optical misalignments than the 1-pair estimates. A space-based coronagraph would not suffer from the same problem because of sturdier mounts, better initial alignment, and greater stability.
  
\subsubsection{Relative Accuracy of the Estimators} \label{sec:accuracy}	

Here we compare the average contrast of the estimated starlight or incoherent light for each estimator. This approach can reveal a net bias but averages out the Gaussian noise of individual pixels. To eliminate variations in images between different correction runs, we ran the estimators on a set of saved images from a 2-pair IEKF correction run. There was no intentionally injected incoherent source. The only difference from using stored data instead of real-time data to feed the estimator is that the control signal was pre-determined, but this did not change the accuracy of the estimator. By using the same images for all estimators, we decoupled correction speed from estimator accuracy. While we cannot know the true state values, we infer that if most of the estimators are close to a value, then it is most likely the true value. 

\begin{figure}[ht!] 
\centering
    \includegraphics[width = .8\columnwidth,trim=.05in 0in .05in 0in]{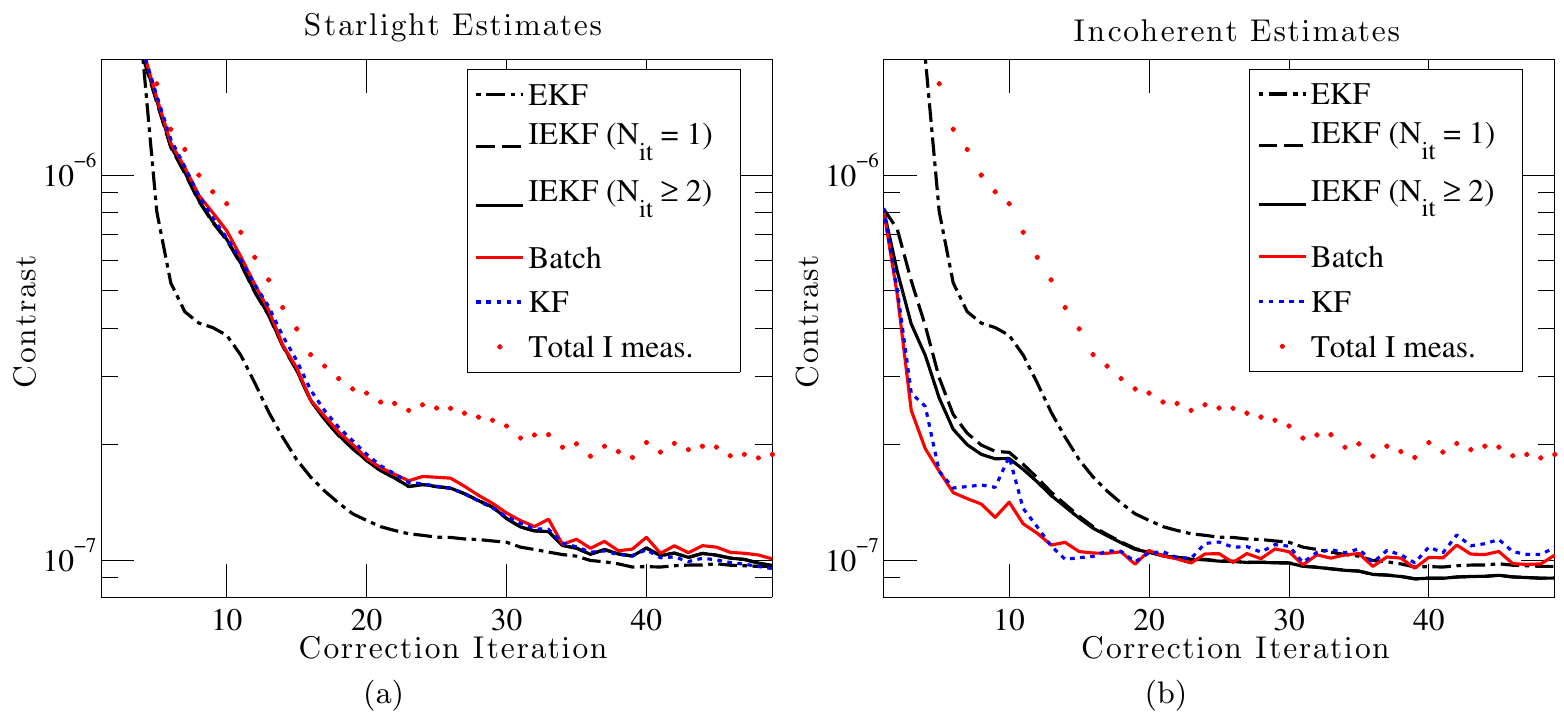}
    \caption{(a) Comparison of the average starlight intensity estimate for each 2-pair estimator. (b) Comparison of the average incoherent intensity estimate for each estimator. $N_{it}$ is the number of EKF iterations performed in the IEKF. All estimators in this case used the same saved set of correction run images to allow a direct comparison of their quality with the exact same data. The measured intensity is included for reference.}
\label{starAndIncoEstCompAll}        
\end{figure}

All the estimators except the EKF gave almost exactly the same average starlight contrast values in Fig.\ \ref{starAndIncoEstCompAll}(a). The EKF exhibited a large bias and mistook much of the starlight for incoherent light, as shown in Fig.\ \ref{starAndIncoEstCompAll}(b). The IEKF eliminated the starlight estimate bias with just one EKF iteration. The batch estimate started to exhibit more fluctuations than the other estimates during the last half of the correction, probably because the batch estimator did not utilize previous estimates to average out read noise. 

The EKF estimates in Fig.\ \ref{starAndIncoEstCompAll} were significantly biased, and yet the contrast correction speed in Fig.\ \ref{estCompAll} clearly shows that the EKF was no slower or less robust at wavefront correction than the IEKF or KF. Somehow a net bias of the estimate at all the pixels did not degrade performance, whereas random errors from noise on batch estimates do slow the correction. This result contradicts our premise that a more accurate estimate yields faster wavefront correction.

\begin{figure}[ht!]   
\centering
    \includegraphics[width = .6\columnwidth,trim=.05in 0in .05in 0in]{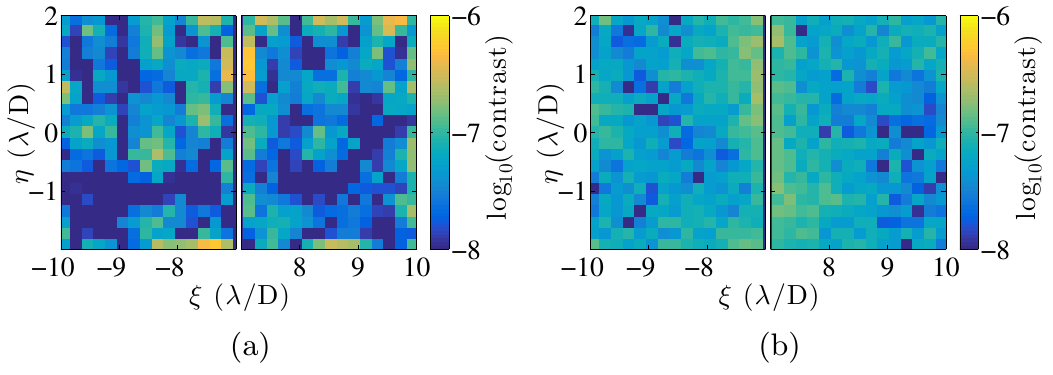}
\caption{Estimates of (a) the starlight intensity at $5{\times}10^{-8}$ average contrast and (b) the incoherent intensity at $6{\times}10^{-8}$ average contrast at the end of the correction run in Fig.\ \ref{fig:typicalRun}. The image plane is cropped to just the dark holes. }
      \label{fig:BoxesOfEstimates}
\end{figure}

The starlight intensity estimate in Fig.\ \ref{starAndIncoEstCompAll}(a) was approximately $1{\times}10^{-7}$ below the measured contrast, and the incoherent estimate converged to this level in Fig.\ \ref{starAndIncoEstCompAll}(b). The incoherent estimate's structure in Fig.\ \ref{fig:BoxesOfEstimates}(b) matched neither that of the coherent estimate in Fig.\ \ref{fig:BoxesOfEstimates}(a) nor that of the nearly flat probe signal, so it was not an artifact of pair-wise estimation. The incoherent signal was also not random, meaning it was not attributable to read noise. We conclude that the incoherent estimate was a true signal composed of stray light.

For the incoherent estimates shown in Fig.\ \ref{starAndIncoEstCompAll}(b), the batch process and KF initially gave more accurate values when the measured contrast was still several times above the true value. In later correction iterations as the contrast curve started to level off, the recursive estimate of the IEKF was much less variable and ended up lower than for the batch process and KF. This suggests that the IEKF adequately filters out read noise. We see in Fig.\ \ref{starAndIncoEstCompAll}(b) that the first EKF iteration helped the most in improving the IEKF incoherent estimate, a second EKF iteration helped a little more in the early iterations, and more than two EKF iterations produced no discernible change. It appears that two EKF iterations are sufficient for convergence of the IEKF estimates. From here on, we stop testing the un-iterated EKF in comparisons because of its heavily biased estimates.


\subsection{Experiments in the Presence of a Flat, Incoherent Background}	

\begin{figure}[ht!] 
\centering
    \includegraphics[width = .8\columnwidth,trim=.05in 0in .05in 0in]{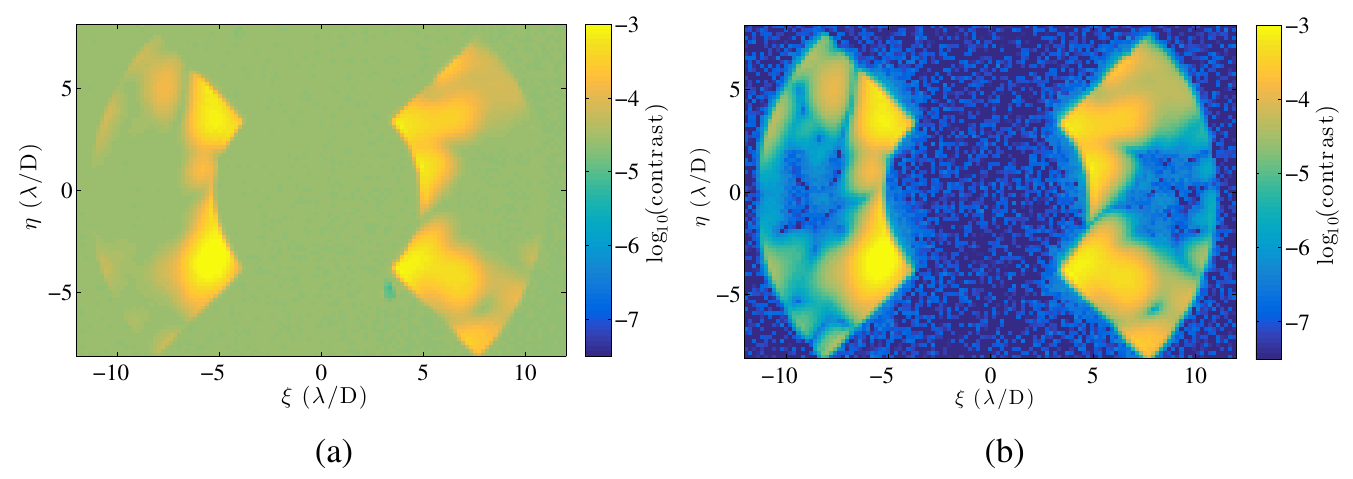}
\caption{Final PSFs for the IEKF correction run when (a) the incoherent background is still on and (b) is temporarily turned off. The incoherent signal appears behind the field stop since it was placed downstream in the system. We used the signal behind the center of the mask  ($\xi \in [-1.5,1.5] \lambda/D$) to determine the net standard deviation at each pixel and the true average incoherent intensity. }
\label{fig_IfinalZodiOnOff_9April2015}
\end{figure}

In this experiment, we compared the 2-pair batch process, KF, and IEKF estimators in the presence of a bright, zodi-like background as shown in Fig.\ \ref{fig_IfinalZodiOnOff_9April2015}(a).  We chose a bright incoherent background at $2.45{\times}10^{-5}$ contrast to give a noise floor worse than the previously achievable contrast floor. The average standard deviation at each pixel was $5.2{\times}10^{-7}$ contrast from readout noise and incoherent-light shot noise, with readout noise alone giving a standard deviation of $8{\times10^{-8}}$. To verify the estimated starlight intensity, we acquired another image with the incoherent source turned off after each correction iteration, as in Fig.\ \ref{fig_IfinalZodiOnOff_9April2015}(b). 

\begin{figure}[ht!] 
\centering
    \includegraphics[width = .8\columnwidth,trim=.05in 0in .05in 0in]{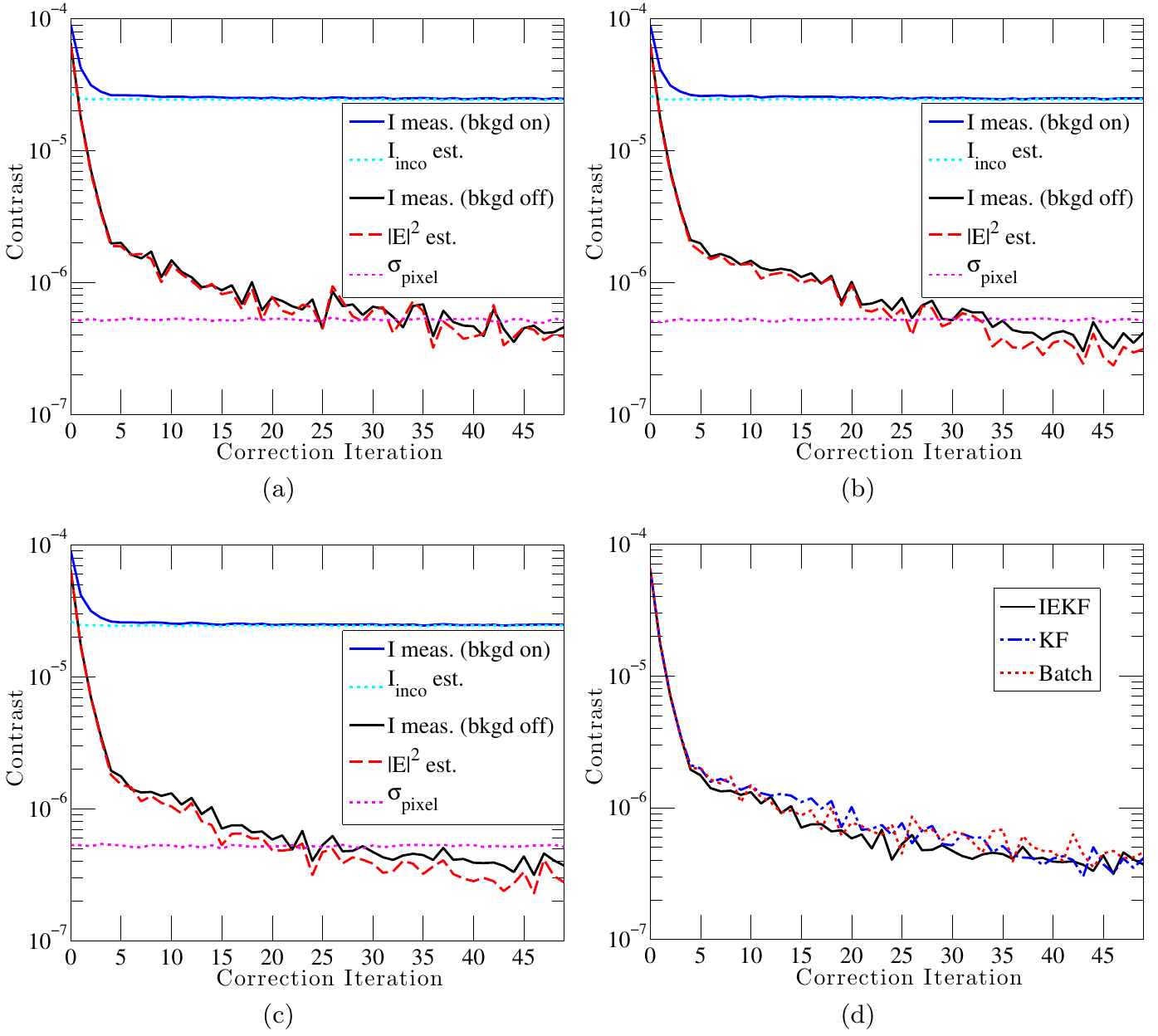}
\caption{Correction runs with different estimators in the presence of $2.45{\times}10^{-5}$ incoherent background light. The mean standard deviation per pixel from zodi shot noise and readout noise was at $5.2{\times}10^{-7}$ contrast. In all measurements, there was always also an incoherent signal from scattered light at about $1{\times}10^{-7}$ average contrast. (a) The batch estimator did well until reaching the noise floor. (b) The KF had slower initial correction speed but allowed correction below the noise floor. (c) The IEKF allowed the fastest correction to the noise floor and below. (d) A direct comparison of the measured intensity with the incoherent background temporarily turned off.}
\label{zodiCcurves}
\end{figure}

We compared the contrast correction speed for the batch process, KF, and IEKF in the presence of the flat background. The nominal incoherent signal at about $1{\times}10^{-7}$ was still present with the zodi turned off, so the estimated starlight intensity should have been below the measured, background-off intensity by that amount. We define the nominal incoherent signal as the stray light when only the starlight laser is on. The KF and IEKF starlight estimates in Figs.\ \ref{zodiCcurves}(b) and \ref{zodiCcurves}(c), respectively, were indeed about $1{\times}10^{-7}$ below the measured, zodi-less intensities, while the batch starlight intensity estimate in Fig.\ \ref{zodiCcurves}(a) was too bright by about $1{\times}10^{-7}$. Directly comparing the measured, zodi-less contrast curves in Fig.\ \ref{zodiCcurves}(d), we did not observe any definitive differences in correction speed or achievable contrast because of the large fluctuations between iterations. 

Compared to the contrast correction curves without zodi in Fig.\ \ref{estCompAll}, the curves with bright background in Fig.\ \ref{zodiCcurves} were much more erratic in their convergence. One possible explanation is that the much larger standard deviation in each measurement was corrupting the estimate quality. Another possibility is that the drifting laser power of the bright background was invalidating the assumption of a static state. Over the course of a correction run, we observed the laser power drift by 1.5\%, corresponding to about $4{\times}10^{-7}$ in contrast. Such a large drift in the allocation of intensity could explain the non-smooth correction curves and the reduced achievable contrast for the bright background case. 
\subsection{Experiments to Recover a Faint Companion}\label{sec:companion}	

\begin{figure}[ht!]   
\centering
    \includegraphics[width = .9\columnwidth,trim=.05in 0in .05in 0in]{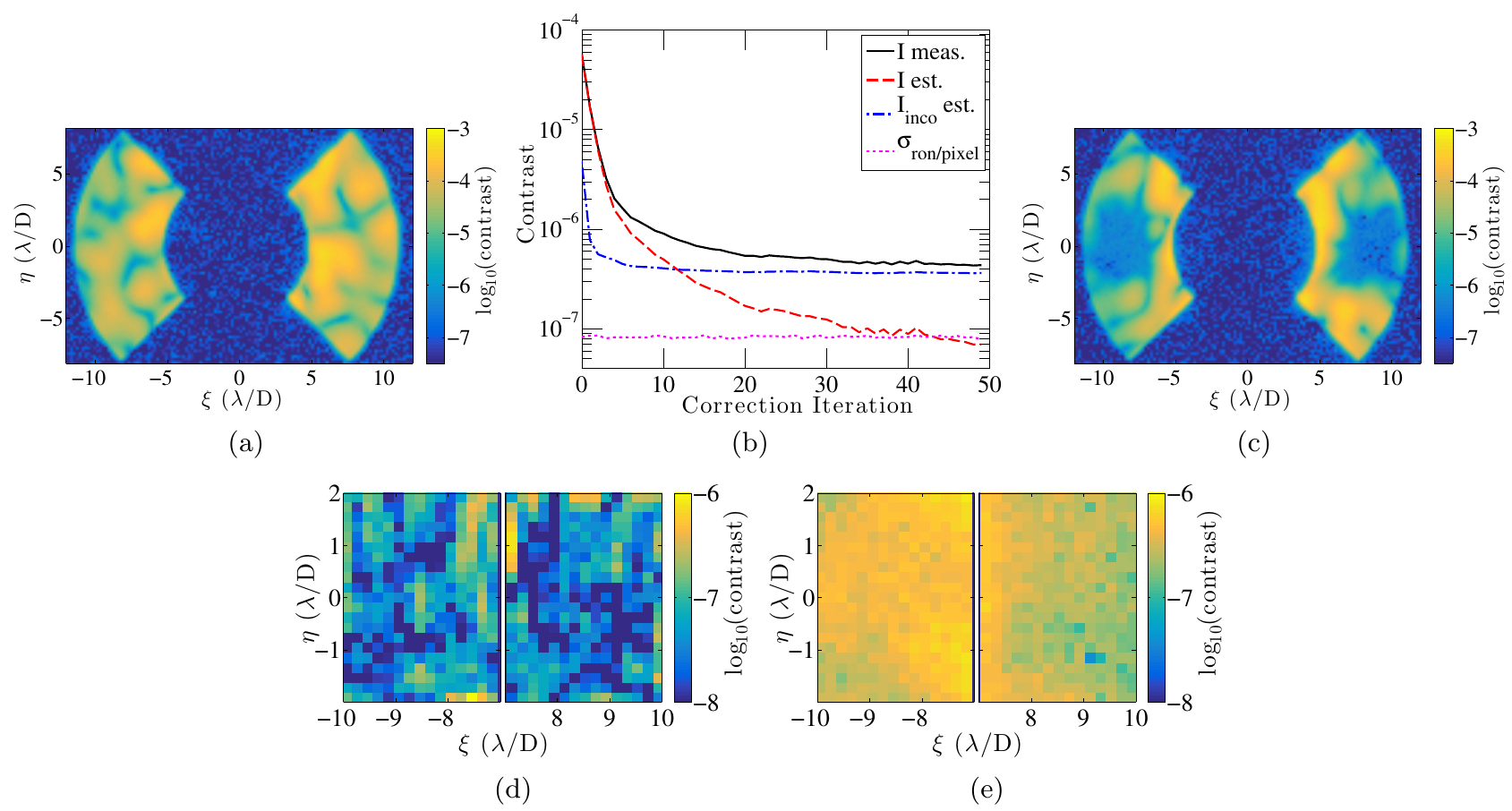}
\caption{Results from an 2-probe-pair IEKF correction run with the star-planet simulator installed but no planet injected. (a) The stopped-down, uncorrected initial image, shown on a log scale, had a contrast of $5.87 \times 10^{-5}$ in the correction region. (c) The final, corrected image had a measured average contrast of $4.4 {\times} 10^{-7}$ in the rectangular dark holes ($3.1 {\times} 10^{-7}$ in just the right-side dark hole). (b) The measured and estimated contrast correction curves. (d) The estimated coherent light after correction. (e) The final estimated incoherent signal, larger from the beamsplitter being placed in the system.}
      \label{fig:noPlanetRun}
\end{figure}

In this set of experiments, we injected a faint, off-axis point source to mimic an exoplanet, and then we attempted to recover its signal. We inserted the star-planet simulator (the beamsplitter and fiber launch 2) into the testbed and used the 2-pair IEKF for each correction run. We first ran wavefront correction without injecting a planet to determine the differences in nominal performance. The initial PSF is shown in Fig.\ \ref{fig:noPlanetRun}(a), and the final, corrected PSF is shown in Fig.\ \ref{fig:noPlanetRun}(c). The estimated starlight correction curve in Fig.\ \ref{fig:noPlanetRun}(b) was approximately as fast as the case without the star-planet simulator in Fig.\ \ref{fig:typicalRun}(b), and the final starlight estimate in Fig.\ \ref{fig:noPlanetRun}(d) was comparable to the one in Fig.\ \ref{fig:BoxesOfEstimates}(a). The beamsplitter introduced a much larger nominal incoherent signal at $3.6 {\times} 10^{-7}$ average contrast as shown in Fig.\ \ref{fig:noPlanetRun}(e).

Next we performed wavefront correction trials with an injected planet at four contrast levels, starting below the average incoherent background level and ending slightly above it. The injected planet used a separate laser channel and was centered at approximately $(\xi,\eta){=}(8.0,-0.6)$. We used the IEKF during real-time correction, saved those images, and re-used them for the KF trials for a more direct estimator comparison. This strategy eliminated variations between trials from noise, hardware, or the controller.  

We compared three different techniques to recover the planet signal from the images. The first was simple PSF subtraction (PS). After a correction run, we took one image with the planet laser on and another with it off. Subtracting these two images yielded the PS estimate. The second technique defined the planet signal as the batch process incoherent estimate (BPIE) from Eq.\ \ref{eq:batchIinco}, where the KF supplied the estimated starlight intensity, $|E_k|^2$. We included the BPIE because it utilized the concept of coherence diversity (i.e., modulating the stellar electric field to distinguish the incoherent signal) without requiring the IEKF developed in this paper. The final method used the recursive incoherent estimate (RIE) from the IEKF as the planet signal. To isolate the planet signal for this analysis, we subtracted the IEKF's best estimate of the nominal incoherent signal, as shown in Fig.\ \ref{fig:noPlanetRun}(e), from the BPIEs and RIEs. Because the incoherent background is unlikely to be fully subtractable in this manner during a space mission, the analysis in this section represents only a best-case scenario. Any non-uniformities or asymmetries in the zodiacal or exozodiacal light would make the background more difficult to subtract.

\begin{figure}[ht!]  
\centering
    \includegraphics[width = .3\columnwidth,trim=.05in 0in .05in 0in]{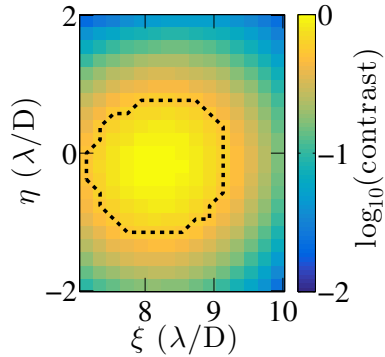}
\caption{Normalized, on-axis PSF shifted to the planet location for use as a PSF template for the planet. Only the region within the full-width at half maximum (shown as the dotted line) was used to avoid fitting to noise.
 }
\label{fig:templatePSF}
\end{figure}

We quantified the quality of the planet signal with two metrics: the accuracy of the planet's contrast estimate and the 2-D correlation of the planet signal to the expected PSF. The contrast estimate was calculated by translating the normalized, on-axis PSF from Fig.\ \ref{fig:SPandPSF}(c) to the planet's location, as shown in Fig.\ \ref{fig:templatePSF},
and then scaling the template PSF's contrast to fit the planet signal. The 2-D correlation, $\mathbb{C}$, which quantitatively compares the morphology of the signals, was calculated by correlating the template PSF, $I_{temp}$, to the extracted planet signal, $\hat{I}_{planet}$,
 \begin{align}
	\mathbb{C} = \frac{ \sum_{s=1}^{N_{FWHM}}\biggl(I_{temp}(s) \hat{I}_{planet}(s)\biggr) }{\sqrt{\biggl(\sum_{s=1}^{N_{FWHM}}I_{temp}(s)^2\biggr) \biggl(\sum_{s=1}^{N_{FWHM}} \hat{I}_{planet}(s)^2 \biggr) }}.
\end{align}
To avoid fitting to noise for both of these metrics, we used only the $N_{FWHM}$ pixels located within the full width at half maximum (FWHM) of the template PSF.

\begin{figure}[ht!] 
\centering
    \includegraphics[width = .5\columnwidth,trim=.05in 0in .05in 0in]{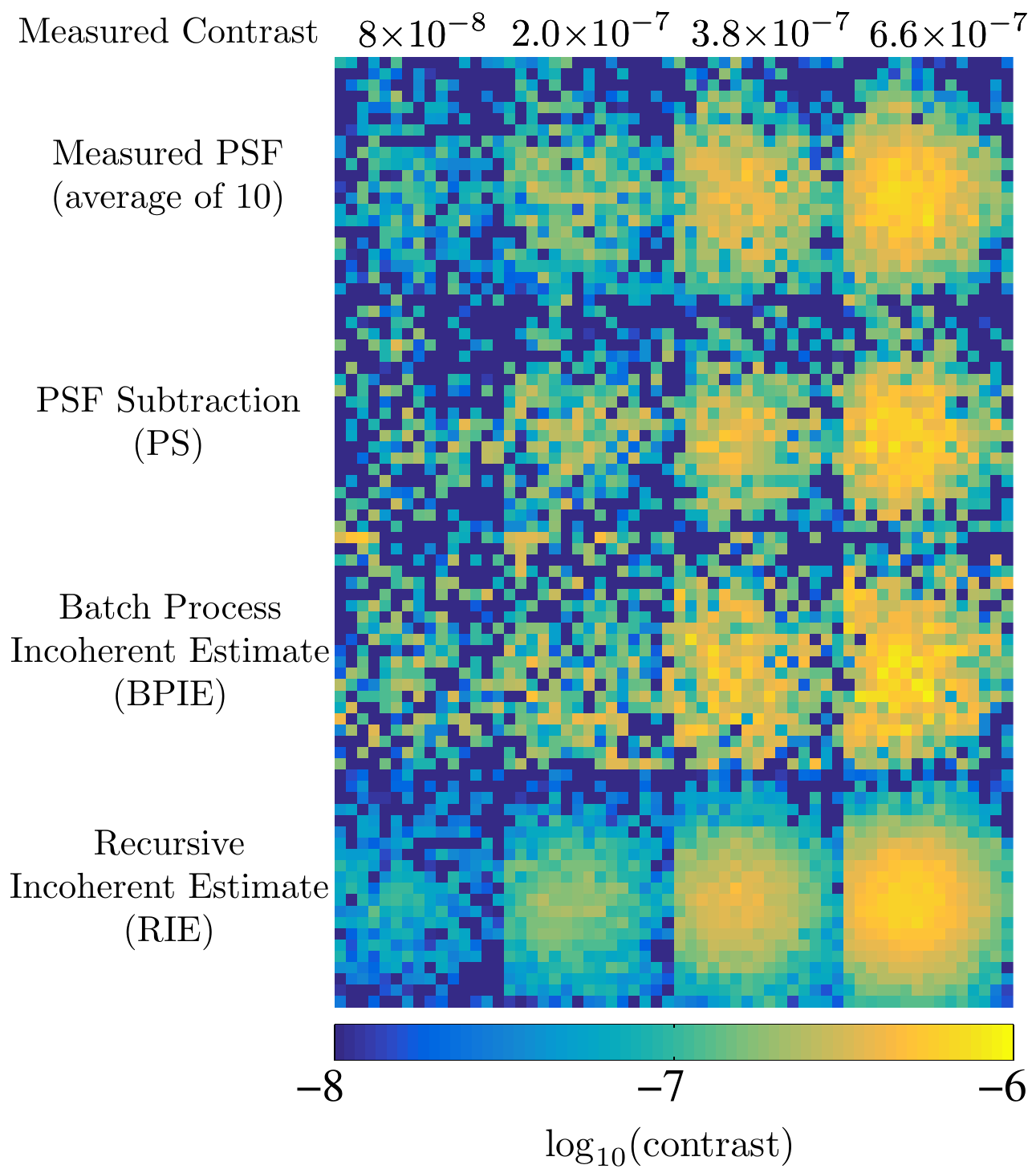}
\caption{Planet signals obtained from the different techniques at the end of correction. Only the right dark hole region is shown.  Each column is for a different planet contrast level. The first row is the planet PSF measured by averaging ten images with the starlight laser off.
 }
\label{fig:gridOfPlanets}
\end{figure}

Figure \ref{fig:gridOfPlanets} shows the planet signal from each routine at each contrast level. The first row displays the planet-only images, which are the averages of ten exposures with the star laser off. The contrast values reported above this row yielded the least-squares fit of the normalized template PSF to the given signal. The average readout noise per pixel in these averaged images was $2.8 {\times} 10^{-8}$ contrast, so the contrast values of $8 {\times} 10^{-8}$, $2.0 {\times} 10^{-7}$, $3.8 {\times} 10^{-7}$, and $6.6 {\times} 10^{-7}$ had signal-to-noise ratios (SNRs) of approximately 2.9, 7.3, 13.9, and 24.1, respectively. The second, third, and fourth rows show the planet signals obtained from the PS, BPIE, and RIE methods, respectively. Upon visual inspection, the RIE produced the least noisy planet signals and the best chance of a detection for the faintest planet setting. 

\begin{figure}[ht!] 
\centering
    \includegraphics[width = .9\columnwidth,trim=.05in 0in .05in 0in]{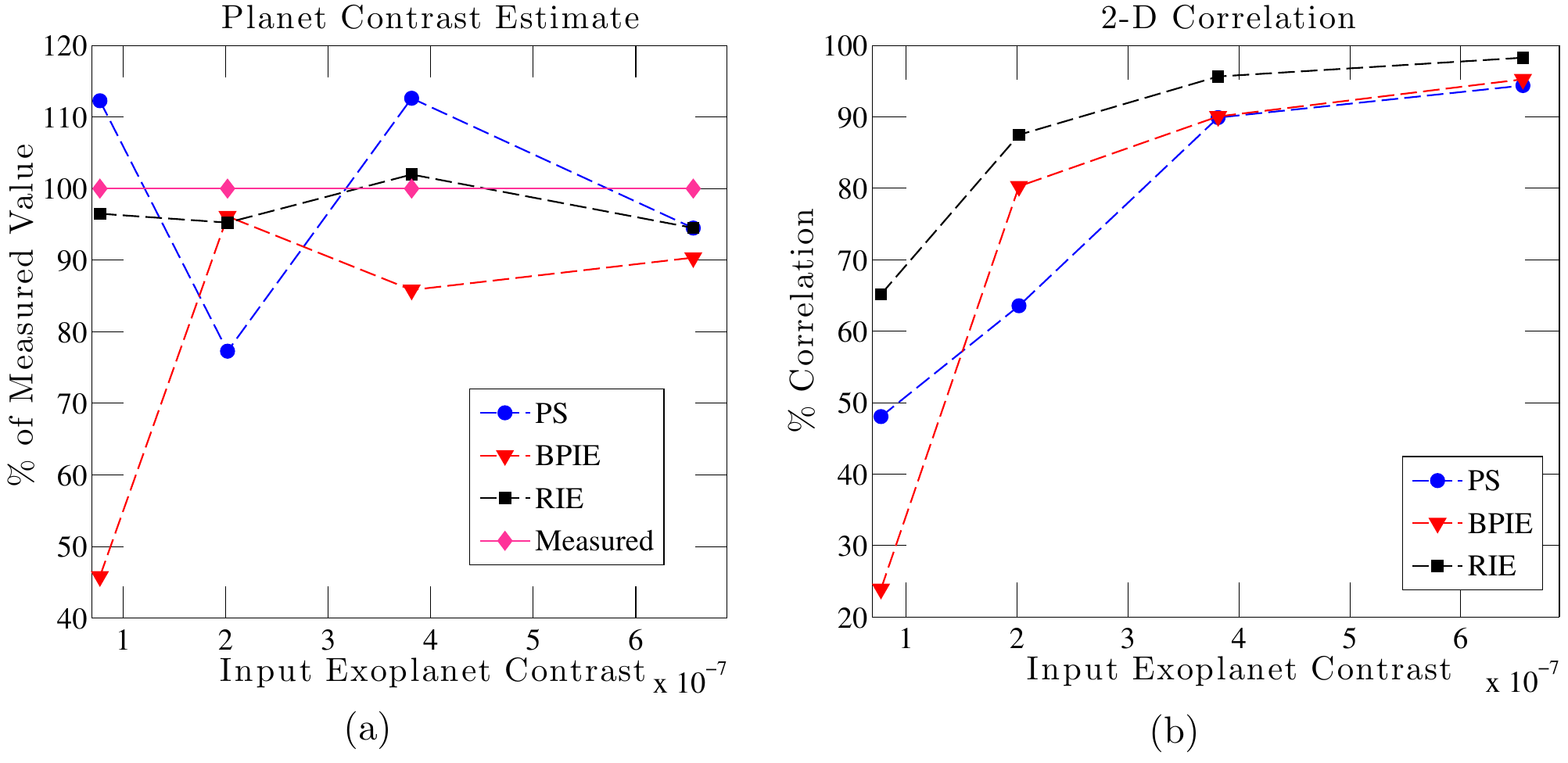}
\caption{(a) Planet contrast estimates and (b) correlation values corresponding to the estimated planet signals in Fig.\ \ref{fig:gridOfPlanets}.
 } \label{fig:endMetrics}
\end{figure}

For the planet signals in Fig.~\ref{fig:gridOfPlanets}, we show the corresponding planet contrast estimates in Fig.~\ref{fig:endMetrics}(a) and correlation values in Fig.~\ref{fig:endMetrics}(b). 
There were not enough data points to determine if PS or the BPIE was more accurate than the other for either metric. The RIE was the only method to be within 5\% of the measured contrast for each planet brightness; higher noise in the other methods yielded much higher error. The correlation of the RIE was always higher than for the other methods at a given planet contrast. The RIE performed well at isolating the incoherent signal, averaging out noise over many iterations, and producing an un-biased contrast estimate. 

The previous analysis used only the final estimates and images from the correction runs. That is the optimal strategy for PSF subtraction, which needs a dimmer dark hole to reduce stellar shot noise in the planet signal, but it might be unnecessary for the BPIE and RIE. Therefore, we calculated the BPIE and RIE after each correction iteration to determine how early they could estimate the planet accurately. We did not take an extra image with the planet laser turned off after each correction iteration, so PS was not included in this comparison. 

\begin{figure}[ht!] 
\centering
    \includegraphics[width = .9\columnwidth,trim=.05in 0in .05in 0in]{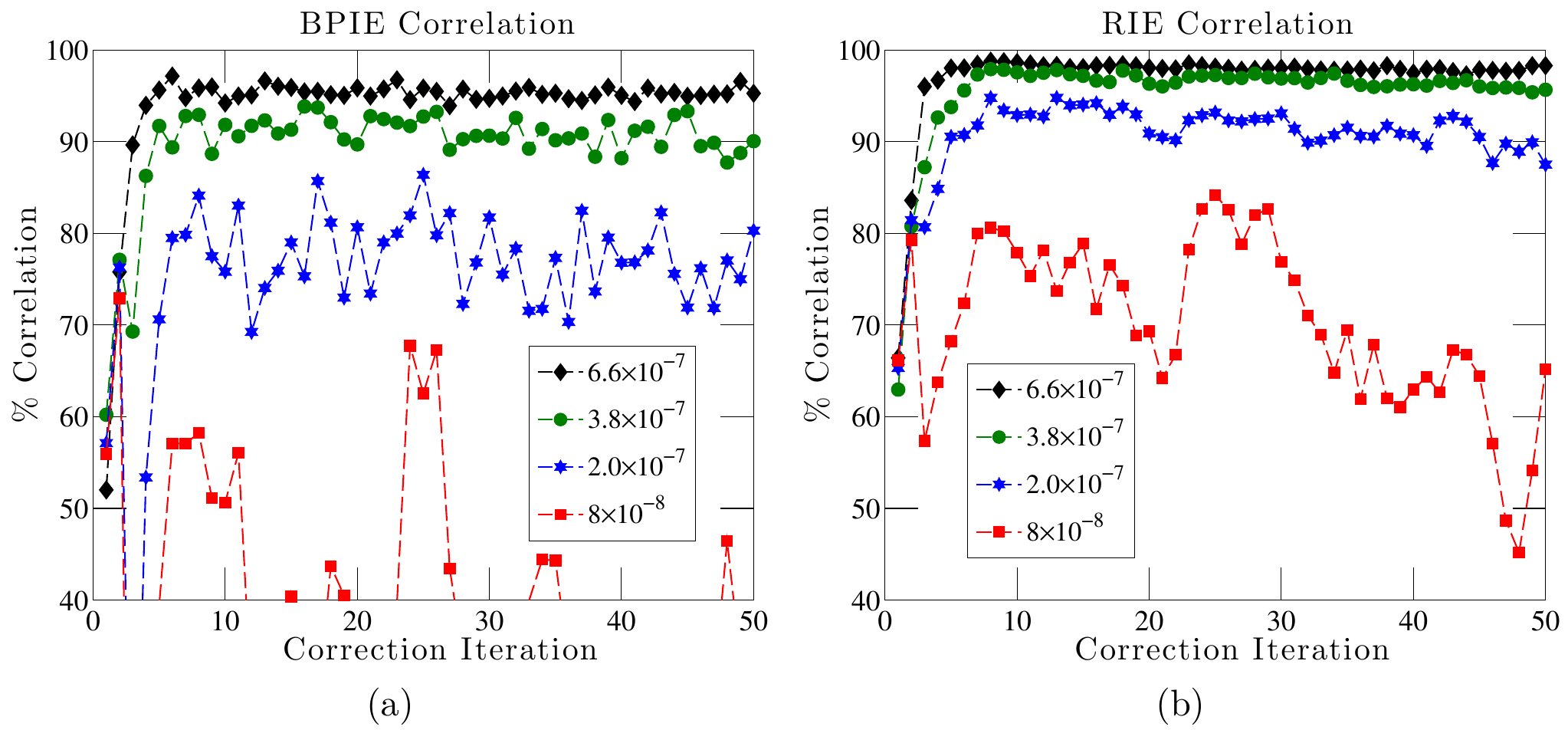}
\caption{Correlation between the planet signal and the template PSF after each correction iteration for (a) the batch process incoherent estimate and (b) the recursive incoherent estimate. Each line corresponds to a different injected planet brightness. Except in the case of the faintest planet, the correlation settles near its final value by about the fifth correction iteration.
 } \label{fig:realtimeCorr}
\end{figure}

Figures \ref{fig:realtimeCorr}(a) and \ref{fig:realtimeCorr}(b) show the BPIE and RIE correlation values, respectively. 
Both methods reached their final values by about the fifth correction iteration, which corresponded to the dark hole reaching approximately $2{\times}10^{-6}$ contrast. The exception was for the faintest planet intensity, for which the correlation values showed much higher variability. For each of the four planet settings, the average RIE correlation was significantly higher and had less variability over time. The faintest two planets merited the most attention as the most difficult ones to detect in a space mission. For the $2.0 {\times} 10^{-7}$ contrast planet in correction iterations 5-50, the mean correlation was 77\% for the BPIE and 92\% for the RIE. Similarly for the faintest planet, the mean correlation was 37\% for the BPIE and 70\% for the RIE. The RIE was thus a much better tool for detecting faint planets than the BPIE.

\begin{figure}[ht!] 
\centering
    \includegraphics[width = .9\columnwidth,trim=.05in 0in .05in 0in]{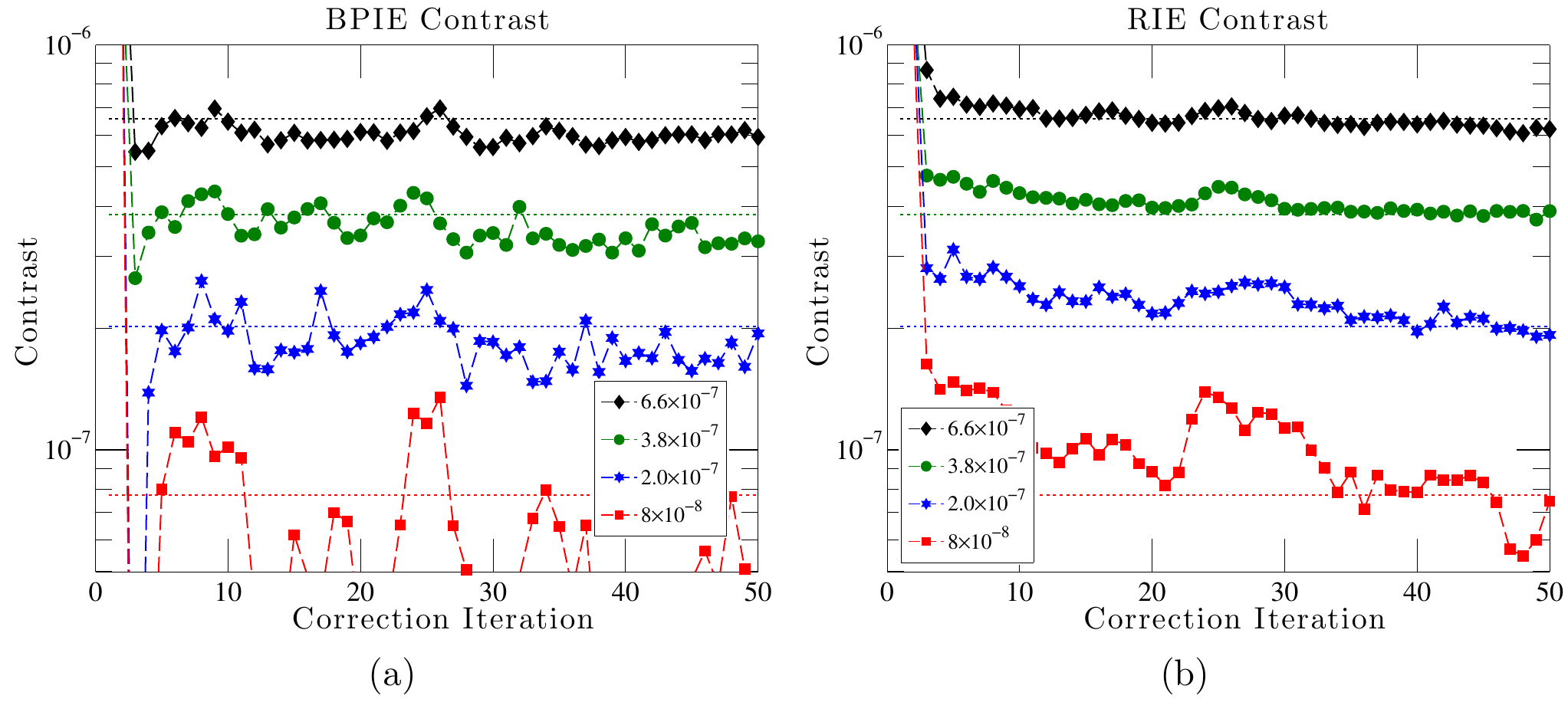}
\caption{Estimated planet contrast after each correction iteration for (a) the batch process incoherent estimate and (b) the recursive incoherent estimate. The measured contrast of the planet is shown for comparison as a dotted line. The BPIE planet contrast values had more variability than the RIE contrasts. The BPIE values settled below the measured contrast levels, while the RIE values started above the measured values before converging to them.
 } \label{fig:realtimeContrastEst}
\end{figure}

Figures \ref{fig:realtimeContrastEst}(a) and \ref{fig:realtimeContrastEst}(b) show the BPIE and RIE planet contrast values, respectively.
The measured contrast levels are shown as dotted lines to reveal biases in the estimates. The values settled in about five correction iterations, and the RIE contrast values showed much less variability among correction iterations compared to the BPIE values. At each planet brightness, the BPIE contrast values started near the measured value but settled with a slightly negative bias. The RIE contrast values started positively biased then settled at the measured contrast. The initial positive bias in the RIE estimates likely arose from starting the IEKF at poor contrast levels, where nonlinearities in the model and observation are large. We previously observed this early-iteration estimate error in Fig.\ \ref{starAndIncoEstCompAll}(b). It may therefore be beneficial to start running the IEKF at moderate contrast levels to avoid the large initial bias in the incoherent estimate.  

We have demonstrated that the incoherent light estimate can be utilized for planet detection during wavefront correction. Because the RIE utilizes the whole history of images to average out noise, it gives the best planet contrast estimate and best PSF correlation compared to PS and the BPIE. These results hold when the other background light can be fully subtracted or is nonexistent, neither of which is safe to assume for a space mission. Non-uniform background light makes planet detection harder for all three of these methods, but these results indicate that the RIE is best at separating starlight from incoherent light. PS may still be the best option if the dark hole speckles are stable long enough to image two different stars. However, if the dark hole does change significantly from slewing the telescope, then coherence diversity via the RIE should be the best option for detecting a companion and estimating its contrast. 

\subsection{Limitations in the HCIL} \label{sec:Limitations} 						

While we have demonstrated the use of an IEKF for generating a dark hole and simultaneously detecting a planet, there are several error sources limiting our attainable contrast and dark hole size.  Scattered light, particularly from the DM mounts, is contributing to a larger than desired background floor.  Our final achievable contrast and the speed at which we reach it are also heavily dependent on our model accuracy.  This is currently limited by our knowledge of the unactuated DM surface, the influence function shapes, and the nominal DM gains. Current work is directed at improving the lab characterization. Finally, the ultimate contrast we can measure is determined by the read noise in the camera, which is high at 4.9 ADU rms for a 40,000 ADU linear range. Future experiments will be performed with a near photon-counting detector. With these changes we expect to reach contrasts close to $10^{-8}$, significantly improving our ability to characterize these algorithms for space missions.

\section{Future Work} \label{sec:futurework}

As we demonstrated by augmenting the state with the incoherent signal, the IEKF also allows the estimation of other system parameters by adding them to the state vector. This is known as parameter adaptive filtering, and it could improve the performance of both the estimator and the controller since both rely on the model. Because the IEKF is sub-optimal from using a linearization of the nonlinear observation, we would like to implement other nonlinear filters such as the unscented Kalman filter to obtain less biased estimates.

In our HCIL wavefront correction trials, we showed that the recursive incoherent estimate provides a higher detection probability and better planet contrast estimate than PSF subtraction. We plan to investigate scenarios that are more representative of a space mission and to quantify the conditions for which either PSF subtraction or the recursive incoherent estimate is better suited. In particular, we would like to simulate trials at higher contrast, with expected levels of speckle dynamics, and at different SNRs. Comparisons assuming a dynamic optical system should also include advanced PSF subtraction techniques such as the Locally Optimized Combination of Images (LOCI)\cite{lafreniere2007loci} and Karhuenen-Lo\`{e}ve Image Projection (KLIP)\cite{soummer2012klip}; Ygouf et al.\ \cite{ygouf2015afta} have already begun this analysis of reference differential imaging for WFIRST-AFTA.    

Up to this point we have used the same set of one unprobed image and $N_{pp}$ probed image pairs per correction iteration because it provides sufficient diversity for wavefront correction. The original purpose of using pairs of probed images was to yield a linear relationship between the electric field and the field change from the probes, but we have just shown that the nonlinear measurement in Eq.\ \ref{eq:ndEKFhofx} along with the IEKF works at least as well if not better. We can therefore modify our measurement equation $z_k$ from Eq.\ \ref{eq:ndEKFz} to a more general one comprised of any set of images. For example, this formulation allows the use of any unpaired probes or even probes on more than a single DM simultaneously. We plan to test other probe combinations in our recursive, nonlinear estimation scheme to further reduce the number of exposures.

\section{Summary}

With the prospect of the WFIRST-AFTA CGI flying in less than a decade, progress in efficient wavefront correction and exoplanet detection algorithms is critical for maximizing the science output of the mission. In our experiments in Princeton's HCIL, we demonstrated the effectiveness of the extended Kalman filter for coronagraphic focal plane wavefront and bias estimation. We found that the EKF and IEKF provide faster wavefront correction by requiring fewer images, and that the IEKF provides a better estimate of the incoherent signal by estimating it recursively along with the starlight. This provides an alternative methodology for separating exoplanet light from the stellar speckles and enables faster, more accurate planet detection. As the simplest nonlinear filters, the EKF and IEKF should be manageable for the computational limits of a space telescope's computer. By proving the viability of using a nonlinear, recursive estimator for focal plane wavefront sensing, we have enabled several possible paths for improvement such as different probe choices and parameter estimation. We found that the IEKF eliminates most of the bias error of the EKF, and we plan to implement more advanced filters for further improvement.

\acknowledgments
A J Eldorado Riggs is funded for this work by the NASA Space Technology Research Fellowship (NNX14AM06H). N.\ Jeremy Kasdin is supported by JPL subaward 1499194 under contract from NASA. Tyler D.\ Groff is funded by the NASA Nancy Grace Roman Technology Fellowship (NNX15AF28G). 


\vspace{1ex}
\noindent{\bf A J Eldorado Riggs} is a Ph.D. candidate in the Department of Mechanical and Aerospace Engineering at Princeton University. He received his BS in physics and mechanical engineering from Yale University in 2011. His research interests include wavefront estimation and control, deformable mirrors, and coronagraph design for exoplanet imaging. He is a member of the American Astronomical Society and SPIE.

\vspace{1ex}
\noindent{\bf N. Jeremy Kasdin} is a Professor of Mechanical and Aerospace Engineering and Vice Dean of the School of Engineering and Applied Science at Princeton University. He is the Principal Investigator of Princeton's High Contrast Imaging Laboratory.  He received his Ph.D. from Stanford University in 1991. Professor Kasdin's research interests include space systems design, space optics and exoplanet imaging, orbital mechanics, guidance and control of space vehicles, optimal estimation, and stochastic process modeling. He is an Associate Fellow of the American Institute of Aeronautics and Astronautics and member of the American Astronomical Society and the SPIE.

\vspace{1ex}
\noindent{\bf Tyler D. Groff} is an associate research scholar at Princeton University. He received his BS in Mechanical Engineering and Astrophysics from Tufts University in 2007, and his Ph.D. from Princeton University in 2012. His current research interests include exoplanet imaging, optomechanics, optical design, wavefront estimation and control, infrared specroscopy, integral field spectrographs, deformable mirrors, and ferrofluidics. He is a member of the American Astronomical Society and SPIE.



\end{spacing}

\end{document}